\documentclass{article}
\usepackage{geometry}
\usepackage{graphicx,epsfig}
\usepackage{natbib}
\usepackage{amsmath,amssymb,amsbsy,mathrsfs}
\usepackage{lscape}
\usepackage{color}
\usepackage{mathtools}

\newcommand{\BluePlus}{\protect \includegraphics{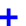}}
\newcommand{\RedCircle}{\protect \includegraphics{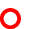}}
\newcommand{\RedTriangle}{\protect
\includegraphics{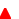}~\protect\includegraphics{marker3}}
\newcommand{\BlueDisc}{\protect
\includegraphics{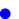}~\protect\includegraphics{marker4}}
\newcommand{\RedDisc}{\protect
\includegraphics{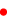}~\protect\includegraphics{marker5}}
\newcommand{\RedSolid}{\protect \includegraphics{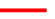}}
\newcommand{\BlueDash}{\protect \includegraphics{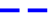}}
\newcommand{\BlackDot}{\protect \includegraphics{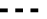}}
\newcommand{\BlackDashDot}{\protect \includegraphics{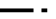}}
\newcommand{\BlueSolid}{\protect \includegraphics{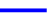}}
\newcommand{\BlackSolid}{\protect \includegraphics{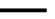}}
\newcommand{\ubar}{{\overline{u}}}
\newcommand{\Bs}{\boldsymbol}

\title{Interaction of linear modulated waves and
unsteady dispersive hydrodynamic states with application to shallow
water waves}

\author{T.~Congy$^1$\thanks{Email address for correspondence:
thibault.congy@northumbria.ac.uk}, G.A.~El$^1$, and
M.A.~Hoefer$^2$}

\date{\normalsize
$^1$Department of Mathematics, Physics and Electrical
Engineering, Northumbria University,
Newcastle upon Tyne NE1 8ST, UK,\\
$^2$ Department of Applied Mathematics, University of Colorado,
Boulder, Colorado 80309-0526, USA}

\begin{document}

\maketitle

\begin{abstract}
A new type of wave-mean flow interaction is identified and studied in
which a small-amplitude, linear, dispersive modulated wave propagates
through an evolving, nonlinear, large-scale fluid state such as an
expansion (rarefaction) wave or a dispersive shock wave (undular
bore).  The Korteweg-de Vries (KdV) equation is considered as a
prototypical example of dynamic wavepacket-mean flow interaction.
Modulation equations are derived for the coupling between linear wave
modulations and a nonlinear mean flow.  These equations admit a
particular class of solutions that describe the transmission or
trapping of a linear wave packet by an unsteady hydrodynamic
state. Two adiabatic invariants of motion are identified that
determine the transmission, trapping conditions and show that
wavepackets incident upon smooth expansion waves or compressive,
rapidly oscillating dispersive shock waves exhibit so-called
hydrodynamic reciprocity recently described in
\cite{maiden_solitonic_2018} in the context of hydrodynamic soliton
tunnelling. The modulation theory results are in excellent agreement
with direct numerical simulations of full KdV dynamics.  The
integrability of the KdV equation is not invoked so these results can
be extended to other nonlinear dispersive fluid mechanic models.
\end{abstract}

\section{Introduction}
\label{sec:intro}

The interaction of waves with a mean flow is a fundamental and
longstanding problem of fluid mechanics with numerous applications in
geophysical fluids (see e.g. \cite{mei_theory_2005},
\cite{buhler_waves_2009} and references therein). Key to the study of
such an interaction is scale separation, whereby the length and time
scales of the waves are much shorter than those of the mean flow.  In
the case of small amplitude, linear waves considered here, the induced
mean flow is negligible so the effectively external mean flow can be
specified separately.  See, for example,
\cite{peregrine_interaction_1976}. The linearised dynamical equations
exhibit variable coefficients due to the mean flow, mathematically
equivalent to the dynamics of linear waves in non-uniform and unsteady
media.

Due to the multi-scale character of wave-mean flow interaction, a
natural mathematical framework for its description is Whitham
modulation theory \citep{whitham_general_1965, whitham_linear_1999}.
Although the initial motivation behind modulation theory was the study
of finite-amplitude waves, it was recognised that the wave action
equation that plays a fundamental role in Whitham theory
\citep{hayes_conservation_1970} was also useful for the study of
linearised waves on a mean flow, see
e.g. \cite{garrett_interaction_1968} and \cite{grimshaw_wave_1984}.
It was used in \cite{bretherton_wavetrains_1968} and
\cite{bretherton_propagation_1968} to examine the interaction between
short-scale, small amplitude internal waves and a mean flow in
inhomogeneous, moving media.  The outcome of this pioneering work was
the determination of the variations of the wavenumber, frequency and
amplitude of the linearised wavetrain along group velocity
lines. Subsequently, this work was extended in
\cite{grimshaw_nonlinear_1975} to finite amplitude waves,
incorporating the perturbative effects of friction and
compressibility, as well as the leading order effect of rotation.

The modulation theory of linear wavetrains in weakly non-homogeneous
and weakly non-stationary media (where weakly is understood as slowly
varying in time and/or space) was developed in
\citep{whitham_general_1965,bretherton_wavetrains_1968}. It was shown
that the modulation system for the wavenumber $k$, the frequency
$\omega$ and the amplitude $a$ is generically composed of the
conservation equations
\begin{equation}
\label{eq:lin_cons}
k_t + \omega_x = 0\,,\quad \mathcal{A}_t+(\partial_k \omega
\mathcal{A})_x=0 \,,
\end{equation}
where the dispersion relation $\omega(k;\Bs \alpha(x,t))$ and the wave
action density $\mathcal{A}(a,k;\Bs \alpha(x,t))$ depend on the system
under study with $\Bs \alpha(x,t)$ being a set of slowly varying
coefficients describing non-homogeneous non-stationary media that
include the effects of the prescribed mean flow, e.g., the current.

Equations~\eqref{eq:lin_cons} were applied to the description of the
interaction of water waves with a steady current in
\citep{longuet-higgins_changes_1961,peregrine_interaction_1976,
phillips_dynamics_1980, peregrine_interaction_1983,
whitham_linear_1999, mei_theory_2005,
buhler_waves_2009,gallet_refraction_2014}. We briefly outline some
classical results from the above references relevant to the
developments in this paper.  Consider a right-propagating surface wave
interacting with a given non-uniform but steady current profile
$U(x)$. Assuming slow dependence of $U$ on $x$, the linear dispersion
relation reads $ \omega = U(x)k + \sigma(k)$ where
$\sigma(k) = \sqrt{g k \tanh(k h)}$ is the so-called intrinsic
frequency, i.e., the frequency of the wave in the reference frame
moving with the current $U$, and $h$ is the unperturbed water
depth. The wave action density has the form
$\mathcal{A} = \rho g a^2/\sigma(k)$.  Since $U$ only depends on $x$,
we look for a steady solution $k(x)$ and $a(x)$ of the modulation
equations~\eqref{eq:lin_cons}, which yield: $\omega_x = 0$ and
$(\partial_k \omega \mathcal{A})_x=0$.  Suppose further that $U(x)$
slowly varies between $U_-=0$ and $U_+$. The wavenumber and the
amplitude of the linear wave then slowly changes from some $k_-,a_-$
to $k_+,a_+$, and the conservation of the frequency and the wave
action yield the following relations:
\begin{equation}
\label{eq:wmfi}
U_+ k_+ + \sigma(k_+) = \sigma(k_-) \,,\quad \frac{U_++\partial_k
\sigma(k_+)}{\sigma(k_+)} \, a_+^2 =
\frac{\partial_k\sigma(k_-)}{\sigma(k_-)} \, a_-^2  \,.
\end{equation}
There is no analytical solution for Eq.~\eqref{eq:wmfi} but one can
check that $k_+$ and $a_+$ are decreasing functions of $U_+$. The
relations~\eqref{eq:wmfi} have been successfully verified
experimentally, cf. for instance \citep{brevik_flume_1980}.  In
particular the linear wave shortens, $k_+>k_-$, and its amplitude
increases, $a_+>a_-$, when it propagates against the current, $U_+<0$.
In this case, the group velocity
$\partial_k \omega(k_+) = U_+ + \partial_k \sigma(k_+)$ vanishes for a
sufficiently short wave, and no energy can propagate against the
current, i.e. the wave is ``stopped'', or ``blocked'' by the current
\citep{taylor_action_1955,lai_laboratory_1989}. Additionally the
amplitude of the linear wave $a_+$ becomes extremely large and the
wave breaks, cf. Eq.~\eqref{eq:wmfi}. As a matter of fact, the linear
approximation fails to be valid for such waves.  As noted in
\citep{peregrine_interaction_1976}, ``such a stopping velocity ...
leads to very rough water surfaces as the wave energy density
increases substantially. Upstream of such points, especially if the
current slackens, the surface of the water is especially smooth as all
short waves are eliminated.'' This phenomenon has been observed when
the sea draws back at the ebb of the tide where an opposing current
increases wave steepness and, as a result, wave breaking occurs (see
for instance \citep{johnson_refraction_1947}).  It also enters in some
pneumatic and hydraulic breakwater scenarios (cf.,
\citep{evans_pneumatic_1955}) where the injection of a local current
destabilises the waves and prevents them from reaching the shore.
More recently, wave blocking has been used to engineer the so-called
white hole horizon for surface waves in the context of analogue
gravity \citep{rousseaux_observation_2008,rousseaux_horizon_2010}.
Finally it has been shown that rogue waves can be triggered when
surface waves propagate against the current in the
ocean~\citep{onorato_triggering_2011}.

Similar problems for a non-uniform and unsteady mean flow have also
been studied. The necessity to consider the media unsteadiness was
first recognised by \cite{unna_white_1941, barber_behaviour_1949,
longuet-higgins_changes_1960}.  Due to the non-stationary character of
the problem, the frequency, as well as the wavenumber, are not
constant. Various unsteady configurations have been studied with
linear theory~\eqref{eq:lin_cons}, such as the influence of an
unsteady gravity constant on water waves
\citep{irvine_kinematics_1985}, the effect of internal waves on
surface waves \citep{hughes_effect_1978}, the water wave-tidal wave
interaction \citep{tolman_influence_1990}, and the influence of
current standing waves on water waves \citep{haller_waves_2007}.

In all described examples, the mean flow or the medium nonhomogeneity
were prescribed externally.  This results in the simple modulation
system~\eqref{eq:lin_cons}, consisting of just two equations with
variable coefficients, with several implications for the wave's
wavelength and amplitude as outlined above.  In this work, we study a
different kind of wave-mean flow interaction, where the mean flow
dynamically evolves in space-time so that the variations of both the
wavetrain and the mean flow are governed by the same nonlinear
dispersive PDE but occur in differing amplitude-frequency domains. The
dynamics of the small amplitude, short-wavelength wave are dominated
by dispersive effects while the large-scale mean flow variation is a
nonlinear process.  In this scenario, the modulation
system~\eqref{eq:lin_cons} for the linear wave couples to an extra
nonlinear evolution equation for the mean flow.  The form of the mean
flow equation depends on the nature of the large-scale unsteady fluid
state involved in the interaction.  For the simplest case of a smooth
expansion (rarefaction) wave, the mean flow equation coincides with
the long-wave, dispersionless, limit of the original dispersive
PDE. However, if the large-scale, nonlinear state is oscillatory, as
happens in a dispersive shock wave (undular bore), the derivation of
the mean flow equation requires full nonlinear modulation analysis
originally presented in \cite{gurevich_nonstationary_1974} (see also
\cite{el_dispersive_2016} and references therein).  We show that in
both cases, the wave-mean flow interaction exhibits two adiabatic
invariants of motion that govern the variations of the wavenumber and
the amplitude in the linear wavetrain, and prescribe its transmission
or {\it trapping} inside the hydrodynamic state: either a rarefaction
wave (RW) or a dispersive shock wave (DSW). Trapping generalises the
aforementioned discussion of blocking phenomena to time-dependent,
nonlinear mean flows.
\begin{figure}
\centering
\includegraphics{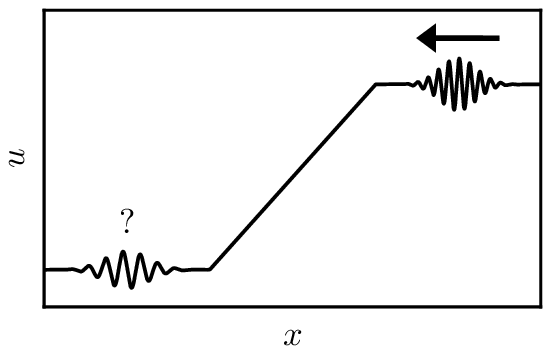}
\includegraphics{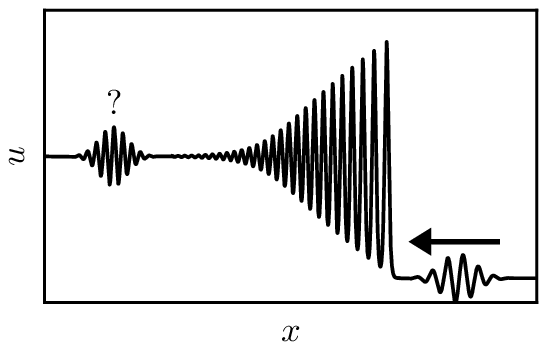}
\caption{\sl Interaction of a linear wavepacket with nonlinear
dispersive hydrodynamic states: a rarefaction wave (left) and a DSW
(right)}
\label{fig:setup}
\end{figure}

As a basic prototypical example, we consider dynamic wavepacket-mean
flow interactions in the framework of the Korteweg-de Vries (KdV)
equation for long shallow water gravity waves:
\begin{equation}
\label{eq:kdv0}
\eta_t + c \eta_x + \frac32 \frac{c}{h} \eta
\eta_x + \frac{h^2 c}{6} \eta_{xxx} = 0 \,,
\end{equation}
where $h$ is the unperturbed water depth, $\eta(x,t)$ is the free
surface elevation relative to $h$ and $c=\sqrt{g h}$ the long-wave
speed.  Equation \eqref{eq:kdv0} exhibits the linear dispersion
relation
\begin{equation}
\omega(k,\overline{\eta}) = c \left ( 1 + \frac{3\overline{\eta}}{2h}
\right ) k - \frac{h^2 c}{6} k^3 \,,
\end{equation}
with frequency $\omega$ and wavenumber $k$.  The KdV equation
describes uni-directional waves that exhibit a balance between weak
nonlinear effects---characterised by the small dimensionless parameter
$\eta_0/h\ll 1$ where $\eta_0$ is the characteristic amplitude of the
free surface displacement---and weak dispersive
effects---characterised by $k_0 h\ll 1$ where $1/k_0$ is a
characteristic horizontal length scale of the perturbation. The
balance leading to the KdV equation is $$\eta_0/h \sim (k_0 h)^2\,,$$
\cite{hammack_modelling_1978}. In particular Eq.~\eqref{eq:kdv0} has
proved effective in the quantitative description of surface waves in
laboratory experiments \citep{zabusky_shallow-water_1971,
hammack_korteweg-vries_1974, hammack_korteweg-vries_1978,
trillo_observation_2016}.

By passing to a reference frame moving at the speed $c$ and
normalising $x$ and $\eta$ by the unperturbed depth $h$, and $t$ by
the characteristic time $h/c$
\begin{equation}\label{phys_rel}
\tilde x = \frac{x - c t}{h},\,
\tilde t = 6 \frac{c t}{h},\,
u = \frac{9 \eta}{h},\,
\end{equation}
the KdV equation Eq.~\eqref{eq:kdv0} assumes its standard form
\begin{equation}
\label{eq:kdv}
u_{\tilde t} + u u_{\tilde x} + u_{\tilde
x \tilde x \tilde x} = 0\,.
\end{equation}
In what follows we shall drop tildes for independent variables and
shall use the normalised equation \eqref{eq:kdv} as our main
mathematical model. All the results obtained in the framework of
Eq.~\eqref{eq:kdv} can then be readily interpreted in terms of the
physical variables using relations \eqref{phys_rel}.  The two basic
settings we consider for Eq.~\eqref{eq:kdv} are illustrated in
Fig.~\ref{fig:setup}. The linear wavepacket propagating with group
velocity $-3 k^2$ relative to the background, say $u=u_0$, is incident
from the right upon an unsteady dispersive-hydrodynamic state: a RW or
a DSW. We derive a system of modulation equations describing the
coupling between the amplitude-frequency modulations in the linear
wave packet and the variations of the background mean flow and show
that the linear wave is either transmitted through or trapped inside
the unsteady hydrodynamic state. The \mbox{transmission/trapping}
conditions are determined by two adiabatic invariants of motion that
coincide with Riemann invariants of the modulation system on a certain
integral surface.

The mathematical approach to the description of dynamic wave-mean flow
interaction that is developed in this paper is general and can be
applied to other models for water waves \citep{lannes_water_2013},
such as Boussinesq type systems describing bidirectional propagation
of nonlinear long waves \citep{bona_boussinesq_2002,
bona_boussinesq_2004, serre_contribution_1953}, the models for short
gravity surface waves \citep{whitham_variational_1967,
trillo_observation_2016}, gravity-capillary waves
\citep{schneider_rigorous_2002}, and others.

The paper is organised as follows.  In Sec.~\ref{sec:mod_dyn}, we
introduce the mean field approximation and linear wave theory to
derive the modulation system for the interaction of a linear modulated
wave with a nonlinear dispersive hydrodynamic state: either a RW or
DSW. This system consists of the two usual modulation equations
\eqref{eq:lin_cons} that describe conservation of wave number and wave
action, which are coupled to the simple wave evolution equation
describing mean flow variations in the RW/DSW. The obtained full
modulation system, despite being non-strictly hyperbolic, is shown to
possess a Riemann invariant associated with the linear group velocity
characteristic.  Moreover, we show that the wave action modulation
equation can be written in diagonal form, effectively exhibiting an
additional Riemann invariant on a certain integral surface.

In Sec.~\ref{sec:scatt_pw}, we consider the model problem of plane
wave-mean flow interaction whereby the mean flow variations are
initiated by Riemann step initial data. Within this framework, the
Riemann invariants of the modulation system found in
Sec.~\ref{sec:mod_dyn} are shown to play the role of adiabatic
invariants of motion that determine the transmission conditions
through the RW. The transmission through a DSW is then determined by
the same conditions as in the RW case by way of hydrodynamic
reciprocity, a notion recently described in the context of
soliton-mean flow interactions \citep{maiden_solitonic_2018}.

The results of Sec.~\ref{sec:scatt_pw} are employed in
Sec.~\ref{sec:numerics} to study the physically relevant case of the
interaction of localised wavepackets with RWs and DSWs. A partial
Riemann problem for wavepacket-RW interaction is used to show that the
variation of the wavepacket's dominant wavenumber is governed by the
conservation of the adiabatic invariant identified in
Sec.~\ref{sec:scatt_pw} and thus yields the same transmission and
trapping conditions for the wave packet as in the full Riemann problem
(plane wave-RW interaction). The same conditions are valid, via
hydrodynamic reciprocity, for the wavepacket-DSW interaction
case. Wavepacket trajectories inside the RW and the DSW are also
determined analytically and compared with the results of numerical
resolution of the corresponding partial Riemann problem. We obtain the
speed and phase shifts of the wavepacket due to its interaction with a
hydrodynamic state.

In Sec.~\ref{sec:conclusions}, we draw conclusions and identify
applications and perspectives for further development of this work.
Appendix~\ref{app:num} describes the numerical implementation of the
partial Riemann problem employed in Sec.~\ref{sec:numerics}.

\section{Modulation dynamics of the linear wave-mean flow interaction}
\label{sec:mod_dyn}

\subsection{Mean field approximation and the modulation equations}
\label{sec:mod_eq}

In this section, we shall introduce the mean field approximation that
enables a straightforward derivation of the modulation system
describing linear wave-mean flow interaction. The full justification
of this approximation for the case of the interaction with a RW can be
done in the framework of standard multiple-scales analysis
\citep{luke_perturbation_1966}, equivalent to single-phase modulation
theory \citep{whitham_linear_1999}. The justification for linear
wave-DSW interaction is more subtle, requiring the derivation of
multiphase (two-phase) nonlinear modulation equations
\cite{ablowitz_evolution_1970} and making linearisation in one of the
oscillatory phases. To avoid unnecessary technicalities, we simply
postulate the approximation used and then justify its validity by
comparison of the obtained results with direct numerical simulations
of the KdV equation.

To describe the interaction of a linear dispersive wave with an
extended nonlinear dispersive-hydrodynamic state (RW or DSW), we
represent the solution $u(x,t)$ of the KdV equation~\eqref{eq:kdv} as
a superposition
\begin{equation}
\label{sol:gen}
u(x,t) = u_{\rm H.S.}(x,t) + \varphi(x,t) \,,
\end{equation}
where $u_{\rm H.S.}(x,t)$ corresponds to the RW or DSW solution, and
$\varphi(x,t)$ corresponds to a small amplitude field describing the
linear wave.

In order to extract the dynamics of $\varphi(x,t)$, we make the mean
field (scale separation) approximation by assuming that
$u_{\rm H.S.}(x,t)$ is locally (i.e. on the scale
$\Delta x \sim \Delta t = O(1)$) periodic and replace the dispersive
hydrodynamic wave field $u_{\rm H.S.}(x,t)$ with its local mean
(period average) value $\ubar(x,t)$.  Within this substitution, the
small amplitude approximation and the mean field assumption read
\begin{equation}
\label{eq:small_slow}
\varphi \ll \ubar \,,\quad \ubar_x/\ubar \ll \varphi_x / \varphi \,,\quad
\ubar_t/\ubar \ll \varphi_t / \varphi.
\end{equation}
For a smooth, slowly varying hydrodynamic state (RW) such a
replacement is natural since locally one has $u_{\rm H.S.}=\ubar$, but
for the oscillatory solutions describing slowly modulated nonlinear
wavetrains in a DSW, $u_{\rm H.S.}(x,t) \ne \ubar$ so the mean field
approximation would require justification via a careful multiple scale
analysis \citep{ablowitz_evolution_1970}. In particular, a detailed
analysis of possible resonances between the DSW and the wavepacket
will be necessary \citep{dobrokhotov_finite-zone_1981}. Such a
mathematical justification will be the subject of a separate work,
while here we shall postulate the outlined mean field approximation
and show that it enables a remarkably accurate description of the
linear field $\varphi$, which can be thought of as propagating on top
of the mean flow.

\begin{figure}
\centering
\includegraphics{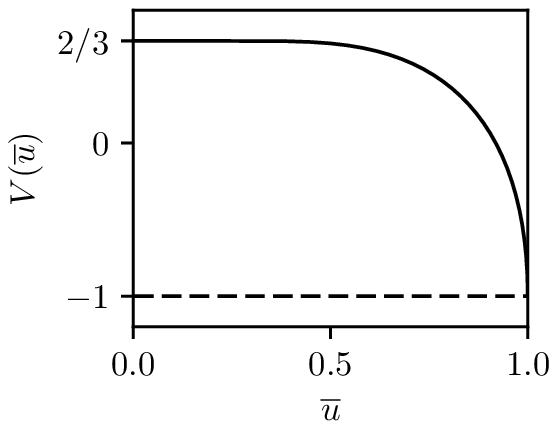}
\includegraphics{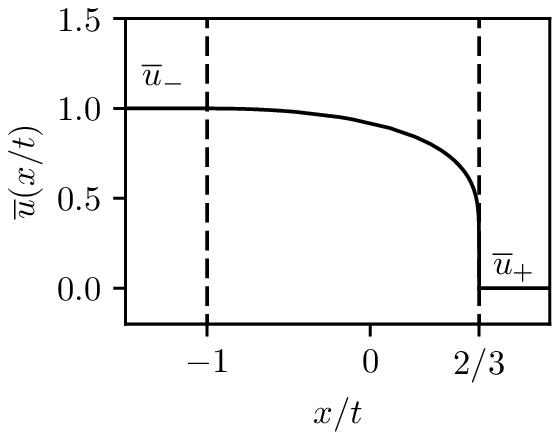}
\caption{\sl Left plot: variation of the characteristic speed of
the Gurevich-Pitaevskii modulation~\eqref{eq:V(u)} with
$(\ubar_-,\ubar_+)=(1,0)$. Right plot: corresponding variation of the
mean flow $\ubar(x/t)$ inside a DSW. The dashed lines correspond to
the DSW edges.}
\label{fig:uz}
\end{figure}

Within the proposed mean field approximation, the small amplitude wave
field $\varphi$ satisfies the linearised, variable coefficient KdV
equation
\begin{equation}
\label{eq:lin_kdv}
\varphi_t + \ubar(x,t) \varphi_x + \varphi_{xxx} = 0\,,
\end{equation}
where the mean flow $\ubar(x,t)$ evolves according to
\begin{equation}
\label{eq:ubar}
\ubar_t + V(\ubar) \,\ubar_x  =0 \,.
\end{equation}
For the case of linear wave-RW interaction: $V(\ubar)=\ubar$, and the
corresponding simplification of Eq.~\eqref{eq:ubar}, known as the Hopf
equation, is obtained by averaging the KdV equation over linear waves
for which $\overline{u^2}=\ubar^2$ and $\overline{u_{xxx}}=0$
\citep{el_resolution_2005}.  For the linear wave-DSW interaction,
$V(\ubar)$ is given parametrically by
\citep{gurevich_nonstationary_1974}
\begin{equation}
\label{eq:V(u)}
\begin{split}
&V= \ubar_+ + \frac13 (\ubar_- - \ubar_+)\left(1+m -
\frac{2m(1-m) K(m)}{E(m)-(1-m)K(m)} \right)
\,, \\
&\ubar = 2\ubar_+-\ubar_- + (\ubar_--\ubar_+)
\left(m +  2 \frac{E(m)}{K(m)} \right) \,,
\end{split}
\end{equation}
where $K(m)$ and $E(m)$ are the complete elliptic integrals of the
first and the second kind respectively
\citep{abramowitz_handbook_1972}, $m \in [0,1]$, and
$\ubar_{-}> \ubar_+$ are the values of $u$ at the left and right
constant states respectively, connected by the DSW.  The parameter $m$
is implicitly obtained as a self-similar solution with $V = x/t$.
Figure \ref{fig:uz} displays the variation of the characteristic speed
$V(\ubar)$ and the mean flow $\ubar$ for
$(\ubar_-,\ubar_+)=(1,0)$. Equations~\eqref{eq:ubar}, \eqref{eq:V(u)}
follow from the Whitham modulation system obtained by averaging the
KdV equation over the family of nonlinear periodic (cnoidal wave) KdV
solutions \citep{whitham_non-linear_1965}. This system consists of
three hyperbolic equations that can be diagonalised in Riemann
invariant form.  The DSW modulation is a simple wave (more
specifically, a 2-wave~\cite{el_dispersive_2016}) solution of the
Whitham equations, in which two of the Riemann invariants are set
constant to provide continuous matching with the external constant
states $u_{\pm}$ \citep{gurevich_nonstationary_1974}, see also
\citep{kamchatnov_nonlinear_2000}.  As we shall show,
Eqs.~\eqref{eq:lin_kdv}, \eqref{eq:ubar}, \eqref{eq:V(u)} provide an
accurate description of the interaction between the linear wave and
the DSW so that the dynamics of $\varphi(x,t)$ are predominantly
governed by the variations of the DSW mean value $\ubar(x,t)$.
We note that a similar mean flow approach, in which the oscillatory
DSW field was replaced by its mean $\ubar$ has been recently
successfully applied to the description of soliton-DSW interaction in
\cite{maiden_solitonic_2018}.

Equations~\eqref{eq:lin_kdv}, \eqref{eq:ubar} form our basic
mathematical model for linear wave-mean flow interaction. We shall
proceed by constructing modulation equations for this system.  One may
question the wisdom of incorporating the decoupled mean flow
equation~\eqref{eq:ubar} to Eq.~\eqref{eq:lin_kdv} rather than simply
prescribing an arbitrary mean flow externally as has been done in
previous works
\citep{bretherton_wavetrains_1968,bretherton_propagation_1968}. As we
will see, the mathematical structure of Eqs.~\eqref{eq:lin_kdv},
\eqref{eq:ubar} enables a convenient solution that is not available
for generic mean flows $\ubar(x,t)$. Moreover,
Eqs.~\eqref{eq:lin_kdv}, \eqref{eq:ubar} transparently reveal the
multiscale structure of the dynamics: a {\it fast}
equation~\eqref{eq:lin_kdv} for the linear waves and a {\it slow}
equation~\eqref{eq:ubar} for the mean flow.

Let $\varphi(x,t)$ describe a slowly varying wavepacket:
\begin{equation}
\label{def:phi}
\varphi(x,t) = a(x,t) \cos\left[ \theta(x,t) \right] \,,\quad \omega =
-\theta_t \,,\quad k = \theta_x \, ,
\end{equation}
where
\begin{equation}
\label{eq:axa}
a_x/a \sim \,k_x/k \sim \omega_x/\omega \ll k \,,\quad
a_t/a \sim k_t/k \sim \omega_t/\omega \ll \omega \, .
\end{equation}
These are standard assumptions in modulation theory
\citep{whitham_non-linear_1965, whitham_linear_1999}, that can be
conveniently formalised by introducing slow space and time variables
$X=\varepsilon t$, $T=\varepsilon t$, where $\varepsilon \ll1$ is a
small parameter, and assuming that $a=a(X, T)$, $k = k(X, T)$,
$\omega = \omega(X,T)$.  To describe the interaction of a linear wave
packet with a nonlinear hydrodynamic state, we require that the slow
variations of the linear wave's parameters and the variations of the
mean flow occur on the same spatiotemporal scale, i.e.
$\ubar = \ubar(X,T)$.  Substituting~\eqref{def:phi}
in~\eqref{eq:small_slow}, we reduce the scale separation conditions
between the linear wave and the mean flow to:
\begin{equation}
\label{eq:small2}
a \ll \ubar\,,\quad \ubar_x/\ubar \ll k\,,\quad \ubar_t/\ubar \ll
\omega \,.
\end{equation}
Conditions ~\eqref{eq:axa} and~\eqref{eq:small2} are the main
assumptions underlying the modulation theory of linear wave-mean
flow interaction described here.  

The derivation of modulation equations for $a$ and $k$ is then
straightforward using Whitham's variational approach (cf. for instance
\citep{whitham_linear_1999}, Ch.~11), and yields
Eq.~\eqref{eq:lin_cons} with:
\begin{equation}
\label{eq:omega}
 \omega = \ubar \,k -k^3,\quad {\cal A} = a^2/k \,.
\end{equation}
We also derive a useful consequence of the wave conservation law
(cf. Eq.~\eqref{eq:lin_cons}) for a {\it wavepacket train} consisting
of a superposition of two slowly modulated plane waves with close
wavenumbers $k$ and $k+\delta k$ where $\delta k \ll k$, which
corresponds to beating of the two waves. The conservation of waves for
these two waves read
\begin{equation}
k_t + \omega(k,\ubar)_x = 0 \,,\quad (k+\delta k)_t + \omega(k+\delta
k,\ubar)_x = 0 \,.
\end{equation}
Hence, for $\delta k \ll k$, the subtraction of these two equations
reduces to a conservation equation for $\delta k$ that is very similar
to the conservation of wave action:
\begin{equation}
\label{eq:delta}
\delta k_t + \left( v_g(k;\ubar) \delta k \right)_x = 0 \,.
\end{equation}

Concluding this section, we note that the modulation
system~\eqref{eq:lin_cons}, \eqref{eq:omega} is quite simple and
definitively not new. However, unlike in previous studies, it is now
coupled to the mean field equation~\eqref{eq:ubar}.  As we shall show,
the system consisting of Eqs.~\eqref{eq:lin_cons}, \eqref{eq:ubar},
\eqref{eq:omega}, and \eqref{eq:delta} equipped with appropriate
initial conditions, yields straightforward yet highly non-trivial
implications, especially in the case of wavepacket-DSW interaction,
which is very difficult to tackle using direct (non-modulation)
analysis.

\subsection{Riemann invariants}
\label{sec:riemann_inv}

In what follows, we shall use an abstract field $A(x,t)$ representing
either ${\cal A}(x,t)=a(x,t)^2/k(x,t)$ or $\delta k(x,t)$ such that
the reduced modulation system composed of Eqs.~\eqref{eq:lin_cons},
\eqref{eq:ubar}, \eqref{eq:omega}, and \eqref{eq:delta} can be cast in
the general form
\begin{subequations}
\label{eq:mod}
\begin{align}
\label{eq:ubar3}
&\ubar_t + V(\ubar) \ubar_x =0\,,\\
\label{eq:k2}
&k_t + v_g(k,\ubar) k_x + \partial_{\ubar} \, \omega(k,\ubar) \,\ubar_x
=0\,,\\
\label{eq:A}
&A_t + \left( v_g(k,\ubar)  A \right)_x = 0\,,
\end{align}
\end{subequations}
where $v_g(k,\ubar) = \partial_k \omega = \ubar -3k^2$,
$\partial_{\ubar} \omega = k$, and $V(\ubar) = \ubar$ or that given in
Eq.~\eqref{eq:V(u)}.  We note that the system~\eqref{eq:mod} has the
double characteristic velocity $v_g$ and thus is not strictly
hyperbolic.  In fact, there are only two linearly independent
characteristic eigenvectors associated with the modulation system
\eqref{eq:mod}, so this system of three equations is only weakly
hyperbolic.  The first two equations, \eqref{eq:ubar3}
and~\eqref{eq:k2}, are decoupled and can always be diagonalised such
that Eq.~\eqref{eq:k2} takes the form
\begin{equation}
\label{eq:q}
q_t + v_g(q,\ubar) q_x = 0 \,,
\end{equation}
for the Riemann invariant $q=Q(k,\ubar)$. Generally, the Riemann
invariant $q$ as a function of $\ubar$ and $k$ is found by integrating
the characteristic differential form
\begin{equation}
\label{diff_form}
\Xi = [\omega_k(k,\ubar)-V(\ubar)] dk + \omega_{\ubar}(k,\ubar) d
\ubar \,.
\end{equation}
For the case of
linear wave-RW interaction, we have $V(\ubar)=\ubar$, so
$\Xi = -3k^2 dk + k d\ubar$ which can be integrated after multiplying
by the integrating factor $1/k$ to yield explicit expressions for the
Riemann invariant $q$ and the associated characteristic velocity
\begin{equation}
\label{eq:qkdv}
q=Q(k,\ubar) = \ubar - \frac32 k^2 \,, \quad  v_g(q,\ubar) = 2q-\ubar\,.
\end{equation}

It follows from~\eqref{eq:q} that $q=\hbox{const} \equiv q_0$ along
the double characteristic $dx/dt=v_g$, which enables one to manipulate
equation~\eqref{eq:A} into the form
\begin{equation}
\label{eq:pA}
(p A)_t + v_g(q_0,\ubar) (p A)_x  = 0\,,
\end{equation}
valid along $dx/dt = v_g$, where
\begin{equation}
\label{eq:p}
p= P(q_0,\ubar)\,, \quad P (q,\ubar)= \exp \left( - \int_{\ubar_0}^{\ubar}
\frac{\partial_u v_g(q,u)}{V(u)-v_g(q,u)} d u
\right) \,,
\end{equation}
with $\ubar_0$ a constant of integration.  The quantity $p A$ thus can
be viewed as a Riemann invariant of the system~\eqref{eq:mod} on the
integral surface $Q(k, \ubar)=q_0$ which will prove useful in the
analysis that follows. We stress, however, that $pA$ is not a Riemann
invariant in the conventional sense since the system~\eqref{eq:mod}
does not have a full set of characteristic eigenvectors. See
\cite{maiden_solitonic_2018} for a similar construction in the context
of soliton-mean flow interaction.

For $V(\ubar)=\ubar$, the integral in~\eqref{eq:p} is readily
evaluated to give, taking into account~\eqref{eq:qkdv},
\begin{equation}
\label{eq:pkdv}
P(q,\ubar) = \sqrt{\frac{\ubar - q}{\ubar_0-q}} = \sqrt{\frac32}
\frac{k}{\sqrt{\ubar_0-q}}\,.
\end{equation}
Note, that the described diagonalisation of the reduced modulation
system~\eqref{eq:mod} for the linear wave-mean flow interaction,
unlike the existence of Riemann invariants of the general Whitham
system for modulated cnoidal waves \citep{whitham_non-linear_1965},
does not rely on integrability of the KdV equation.  In fact, the
possibility of this diagonalisation is general and is a direct
consequence of the absence of an induced mean flow for linearised
waves, so that the dynamics of the wave parameters $(k,\omega,a)$ are
decoupled from the dynamics of the mean flow $\ubar$.  Here, we reap
the benefits of jointly considering the evolution of the mean flow,
wavenumber conservation, and the field equation in~\eqref{eq:mod} by
recognising that they can be cast in diagonal, Riemann invariant form
\eqref{eq:ubar3}, \eqref{eq:q}, and~\eqref{eq:pA} along
$Q(k,\ubar)=q_0$.

\section{Plane wave-mean flow interaction: the generalised Riemann
problem}
\label{sec:scatt_pw}

\subsection{Adiabatic invariants and transmission conditions}

\begin{figure}
\centering
\includegraphics{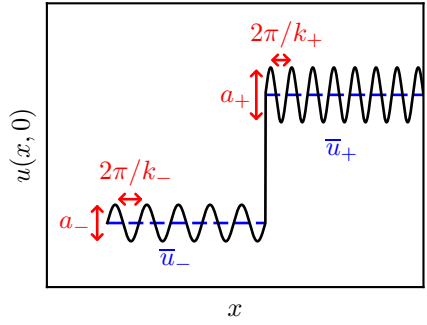}
\caption{\sl Schematic of the initial conditions for the interaction
between a plane wave and a hydrodynamic state.}
\label{fig:pw_scatt_scematic}
\end{figure}

Before studying the interaction of localised wavepackets with a mean
flow in the framework of the basic system~\eqref{eq:lin_kdv},
\eqref{eq:ubar}, we consider a model problem of the unidirectional
scattering of a linear plane wave (PW) by a nonlinear hydrodynamic
state (RW or DSW) initiated by a step in $\ubar$.  We denote the
incident PW parameters at $x \to +\infty$ as $k_+, a_+$ and the
transmitted PW parameters at $x \to -\infty$ as $k_-, a_-$. To find
the transmission relations, we consider the generalised Riemann
problem (see Fig.~\ref{fig:pw_scatt_scematic})
\begin{equation}
\label{eq:t0}
\ubar(x,0),\,k(x,0),\,a(x,0) =
\begin{cases}
\ubar_-,k_-, a_- &\text{if }x<0 \\
\ubar_+,k_+,a_+ &\text{if }x>0
\end{cases}\,.
\end{equation}
We call this Riemann problem {\it generalised} as it is formulated for
the modulation system~\eqref{eq:mod} rather than for the original
dispersive model \eqref{eq:lin_kdv}, \eqref{eq:ubar}.

In the interaction of a PW with both a RW and a DSW, the
evolution of $\ubar(x,t)$ is described by the self-similar expansion
fan solution of the mean flow equation~\eqref{eq:ubar3}:
\begin{equation}
\label{sol:ubar}
\ubar(x,t) =
\begin{cases}
\ubar_- &\text{if} \quad x/t < V(\ubar_-) \\
V^{-1}(x/t) &\text{if} \quad V(\ubar_-) \leq x/t < V(\ubar_+)\\
\ubar_+ &\text{if} \quad x/t  \geq V(\ubar_+)\end{cases},
\end{equation}
while the Riemann invariants $q$ and $pA$ are constant throughout,
implying the relations
\begin{align}
\label{sol:q}
&q=Q(k,\ubar) = Q(k_-,\ubar_-) = Q(k_+,\ubar_+) \,,\\
\label{sol:pA}
&p A=P(k,\ubar) A = P(k_-,\ubar_-) A_-= P(k_+,\ubar_+) A_+\,,
\end{align}
where $A_\pm = a^2_\pm/k_\pm$.  These conserved quantities generalise
the conservation of wave frequency and wave action,
Eq.~\eqref{eq:wmfi}, for steady mean flows to the unsteady case.  The
expressions for $Q(k, \ubar)$ and $P(k, \ubar)$ in the PW-RW
interactions for KdV are given by Eqs.~\eqref{eq:qkdv}
and~\eqref{eq:pkdv}, respectively. For PW-DSW interaction, when
$V(\ubar)$ is given by Eq.~\eqref{eq:V(u)}, simple explicit
expressions for $Q$ and $P$ in terms of $k$ and $\ubar$ are not
available. However, they can be obtained by
integrating~\eqref{diff_form} and evaluating~\eqref{eq:p}, e.g.,
numerically. The edge speeds of the expansion fan~\eqref{sol:ubar} are
given by
\begin{eqnarray}
&&\hbox{RW}:  \  \ \qquad V(\ubar_\pm) =
\ubar_{\pm}   \label{edges_RW} \,,\\
&& \hbox{DSW}: \qquad   V(\ubar_-)=  2 \ubar_+- \ubar_-, \quad
V(\ubar_+)=  \frac{1}{3}(\ubar_{+} + 2 \ubar_-)  \label{edges_DSW} \,.
\end{eqnarray}
Expressions~\eqref{edges_DSW} follow from Eq.~\eqref{eq:V(u)} upon
taking $m \to 0^+$ and $m \to 1^-$ for the trailing and the leading
DSW edge respectively, see \cite{gurevich_nonstationary_1974}.

Given $\ubar(x,t)$ described by~\eqref{sol:ubar}, the conservation of
$q$ and $p A$ in~\eqref{sol:q}, \eqref{sol:pA} yields not only the PW
transmission relations but also the slow variations of the PW
parameters $k(x/t)$ and $a(x/t)$ due to the interaction with the mean
flow in the hydrodynamic state.  Constant $q$ and $p A$ can thus be
seen as {\it adiabatic invariants} of the PW-mean flow interaction.

The conservation relations~\eqref{sol:q}, \eqref{sol:pA} also describe
wave-mean flow interaction for a system of two beating, superposed
slowly modulated plane waves with close wavenumbers $k$ and
$k+\delta k$ interacting with a RW (cf. Sec. \ref{sec:mod_eq}). In
this case, the adiabatic variation of $k$ and $\delta k$ are described
respectively by~\eqref{sol:q} and~\eqref{sol:pA} where $A$ corresponds
now to $\delta k$.

As we already mentioned, relations~\eqref{sol:ubar}, \eqref{sol:q} and
\eqref{sol:pA} are valid for both PW-RW ($\ubar_-<\ubar_+$) and PW-DSW
($\ubar_- > \ubar_+$) interactions.  We now consider these two cases
in more detail.

\subsection{Plane wave-rarefaction wave interaction}
\label{sec:adiabatic_rw}

A RW is generated when $\ubar_-<\ubar_+$, and the resulting mean flow
variation is described by Eq.~\eqref{sol:ubar} with characteristic
velocity $V(\ubar)=\ubar$. In this case, explicit expressions for the
adiabatic invariants $q$ and $pA$ can be obtained using
Eqs.~\eqref{eq:qkdv} and~\eqref{eq:pkdv}. The conservation relations
\eqref{sol:q}, \eqref{sol:pA} then yield
\begin{align}
\label{sol:k}
&\ubar_- - \frac32 k^2_- = \ubar_+ - \frac32 k^2_+\,, \\
\label{sol:a}
&a_- = a_+\,,
\end{align}
where the second condition was obtained by using
$A_{\pm}=a_{\pm}^2/k_\pm$. It is surprising that the interaction of a
PW with a non-uniform, unsteady hydrodynamic state does not change the
PW amplitude, which is in sharp contrast with the classical case of
the interaction of a surface water wave with a counter-propagating
steady current where the amplitude varies following the
inhomogeneities of the current in Eq.~\eqref{eq:wmfi}. In this latter
case, the wave amplitude can become extremely large during the
interaction with the mean flow (and hence the wave is no longer
described by linear theory) while Eq.~\eqref{sol:a} describing dynamic
wave-mean flow interaction ensures that the PW remains a
small-amplitude linear wave regardless of its wavenumber. In
particular no wave-breaking occurs during the interaction with a RW.

Fig.~\ref{fig:k} displays the comparison between the
relation~\eqref{sol:k} and the wavenumbers obtained in numerical
simulations of linear wave-mean flow interaction. In numerical
simulations, we employed the more adequate partial Riemann problem
defined in Sec.~\ref{sec:partial} for which we will show that the
relation~\eqref{sol:k} remains valid. One can see that
Eq.~\eqref{sol:k} yields the transmission condition: the transmitted
PW exists if its wavenumber $k_-$ is a real number. This requirement
reduces to the following condition for transmission of the incident PW
with wavenumber $k_+$:
\begin{equation}
\label{sol:trans}
k_+ > k_c=\sqrt{\frac23 |\ubar_+- \ubar_-|}\,.
\end{equation}
The case $k_+ < k_c$ will receive further interpretation in
Sec.~\ref{sec:numerics} in the context of the interaction of a
localised wavepacket with a RW as wave trapping inside the
hydrodynamic state.

One can also consider the interaction between two beating superposed
PWs and a RW (see the discussion in Sec.~\ref{sec:mod_eq}) where the
conservation of the adiabatic invariant $p(k,\ubar)A$ with
$A = \delta k$ yields:
\begin{equation}
\label{sol:deltak}
k_- \delta k_- = k_+ \delta k_+ \,.
\end{equation}
As was mentioned in the previous section, the beating pattern created
by the superposition of two PWs with close wavenumbers
($\delta k \ll k$) can be seen as a wavepacket train of period
$L = 4\pi/\delta k$.  Thus Eq.~\eqref{sol:deltak} provides the
relation between the wavelength of the incident train $L_+$ and the
transmitted train $L_-$. The difference
\begin{equation}
\label{sol:Delta}
\Delta_- =  L_- - L_+ =  L_+ \left (\frac{k_-}{k_+}-1\right)
\end{equation}
can be interpreted as the phase shift between the incident and the
transmitted wavepackets.
Similarly, one can interpret $L_\pm$ as the widths
of the wavepackets before and after transmission. Their relation is
then given by
\begin{equation}
\label{sol:width}
\frac{k_-}{L_-} = \frac{k_+}{L_+} \,.
\end{equation}

\subsection{Plane wave-DSW interaction: hydrodynamic reciprocity}
\label{sec:adiabatic_dsw}

We now consider the initial condition~\eqref{eq:t0} with
$\ubar_- > \ubar_+$ that resolves into a DSW. In this case, the
modulation of the mean flow is described by the simple wave equation
~\eqref{eq:V(u)}, \eqref{sol:ubar} and the expressions for the
adiabatic invariants $q$ and $pA$ differ from Eqs.~\eqref{eq:qkdv},
\eqref{eq:pkdv} obtained for PW-RW interaction. As a result, the
conditions~\eqref{sol:q}, \eqref{sol:pA} for the conservation of $q$
and $pA$ describe a very different adiabatic evolution of the PW
parameters inside the dispersive hydrodynamic state. We shall consider
this evolution later in Sec.~\ref{sec:wp-dsw}, while here we describe
a very general property of PW interaction with dispersive hydrodynamic
states termed {\it hydrodynamic reciprocity}, that was initially
formulated in \cite{maiden_solitonic_2018} for mean field interaction
of solitons with dispersive hydrodynamic states.

\begin{figure}
\centering
\includegraphics{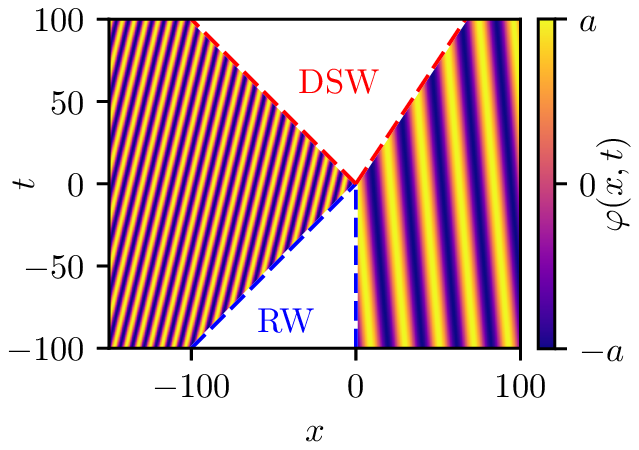}
\includegraphics{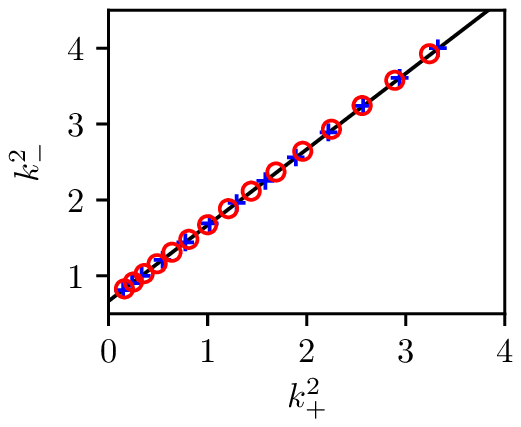}
\caption{\sl Hydrodynamic reciprocity of plane wave-RW and plane
wave-DSW interactions: conservation of $Q(k,\ubar)$ and $a$ in
Eqs.~\eqref{sol:k} and \eqref{sol:a}.  Left plot: modulated plane
wave $\varphi = a \cos{\theta}$, $\theta_x = k$,
$\theta_t = - \omega(k,\ubar)$, outside of the domain of interaction
with the hydrodynamic state, where $\ubar_->\ubar_+$, describing
PW-DSW interaction for $t>0$ and PW-RW interaction for $t<0$. Right
plot: the relationship between the dominant wavenumbers of incident
and transmitted wavepackets with $a_0 = 0.01$.  The solid line
corresponds to the analytical relation~\eqref{sol:k} when
$(\ubar_-,\ubar_+)=(1,0)$. The crosses (\BluePlus) and circles
(\RedCircle) are identified with PW-RW and PW-DSW interaction,
respectively.  The relation between $k_-$ and $k_+$ is independent
of the nature of the hydrodynamic state.}
\label{fig:k}
\end{figure}

When $\ubar_- > \ubar_+$, we observe that the PW-DSW and PW-RW
interactions in the mean flow approximation are described by the
solutions of the same Riemann problem considered for the $t>0$ and
$t<0$ half-planes, respectively. Then, continuity of the simple wave
modulation solution for all $(x,t)$ (illustrated by Fig.~\ref{fig:k},
left), except at the origin $(x,t)=(0,0)$, implies that the transition
relations~\eqref{sol:k} and~\eqref{sol:a} derived for the PW-RW
interaction ($t<0$) must also hold for $t>0$, i.e., for the PW-DSW
interaction. This hydrodynamic reciprocity is verified in
Fig.~\ref{fig:k}, right where we compare the relations between $k_-$
and $k_+$ obtained numerically for the evolution of PW-RW and PW-DSW
interactions in the full KdV equation. The agreement confirms that
relation~\eqref{sol:k}, as well as the transmission
condition~\eqref{sol:trans}, indeed hold for PW interaction with both
nonlinear dispersive hydrodynamic states: RW and DSW.

The agreement between relation~\eqref{sol:k} and the numerical
solution of the Riemann problem in Fig.~\ref{fig:k} also confirms the
mean field hypothesis underlying the basic mathematical
model~\eqref{eq:lin_kdv} of this paper.  Although the mean field
assumption is arguably intuitive for PW-RW interaction where the
hydrodynamic state solution $u_{\rm H.S.}$, up to small dispersive
corrections at the RW corners, coincides with the solution of
\eqref{eq:ubar} for mean flow evolution $\ubar=V(\ubar)$, it is no
longer so for the highly nontrivial PW-DSW interaction where
$u_{\rm H.S.}$ describes a rapidly oscillating structure, which is
radically different from its slowly varying mean flow $\ubar$
satisfying the equations~\eqref{eq:ubar}, \eqref{eq:V(u)}.

\begin{figure}
\centering
\includegraphics{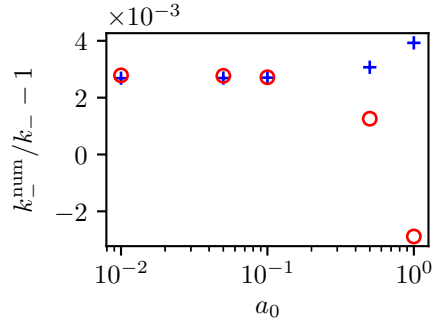}
\caption{Relative error $k_-^{\rm num}/k_--1$ where $k_-^{\rm num}$ is
obtained numerically and $k_-$ satisfies \eqref{sol:k} with
$k_+=0.88$ and $(\ubar_-,\ubar_+) = (0,1)$ (PW-RW) or
$(\ubar_-,\ubar_+) = (1,0)$ (PW-DSW).  The numerical results are
obtained for different amplitudes $a_0$ of the incident wavepacket.
The crosses (\BluePlus) and circles (\RedCircle) correspond to the
interaction with a RW and DSW, respectively.}
\label{fig:delta_k}
\end{figure}

The relative difference between the numerically observed wavenumber
$k_-^{\rm num}$ and the predicted wavenumber $k_-$ for transmitted
wavepackets through RWs and DSWs is shown in Fig.~\ref{fig:delta_k}.
While the relative error in the small amplitude regime reported in
Fig.~\ref{fig:k} is on the order of $10^{-3}$, it is surprising that
the wavenumber prediction from linear theory holds equally as well for
wavepackets with order one amplitudes.

\subsection{Nonlinear plane wave-mean flow interaction}
\label{sec:nonlinear-plane-wave}

It is worth taking a brief detour from our general approach, which is
applicable to a broad class of nonlinear, dispersive equations, to
focus on the KdV equation itself.  The reason for this is that, for
waves governed by the KdV equation, we can fully describe nonlinear
plane wave interaction utilising Whitham theory.  We discover three
intriguing facts: 1) an arbitrarily large, transmitted nonlinear plane
wave conserves its amplitude when interacting with a mean flow (RW or
DSW), 2) the linear plane-wave transmission condition in
Eq.~\eqref{sol:k} accurately describes nonlinear plane-wave
transmission, and 3) the induced mean flow due to the nonlinear wave
is negligible.

For this, we introduce the KdV-Whitham equations that describe the
slow modulations of a nonlinear periodic travelling wave solution of
the KdV equation \eqref{eq:kdv}
\citep{whitham_non-linear_1965,el_dispersive_2016}
\begin{equation}
\label{eq:1}
\frac{\partial r_j}{\partial t} + V_j \frac{\partial r_j}{\partial
x} = 0\,, \quad j = 1,2,3 \,,
\end{equation}
where $r_1$, $r_2$, and $r_3$ are the modulation parameters that vary
slowly relative to the nonlinear wave's wavelength $2\pi/k$ and
temporal period $2\pi/\omega$.  A remarkable feature of the
KdV-Whitham equations---owing to KdV's integrable structure---is that
$r_1$, $r_2$, and $r_3$ are all Riemann invariants.  The
characteristic velocities $V_j$ are known, ordered
($V_1 \le V_2 \le V_3$), nonlinear functions of the modulation
parameters but we will only require
\begin{equation}
\label{eq:4}
V_3 =  \frac{1}{3}(r_1+r_2+r_3) + \frac{2}{3}(r_3-r_1) \frac{(1-m)
K(m)}{E(m)} \,,  \quad m = \frac{r_2 - r_1}{r_3 - r_1} \,.
\end{equation}
Note that $V_2$ is given in Eq.~\eqref{eq:V(u)} (with $r_1 = 0$,
$r_2 = m$, and $r_3 = 1$) and determines the DSW mean flow variation.
The relationship between the parameters $r_j$ and KdV's nonlinear
periodic travelling wave solution is
\begin{equation}
\label{eq:2}
\varphi(x,t) = r_1 + r_2 - r_3 + 2(r_3 - r_1) \mathrm{dn}^2 \left (
\frac{K(m)}{\pi} \theta; m \right )\,,
\end{equation}
where $\mathrm{dn}$ is a Jacobi elliptic function and
$\theta = \theta(x,t)$ is the wave's phase that satisfies the
generalised frequency and wavenumber conditions $\theta_t = - \omega$,
$\theta_x = k$ (cf.~Eq.~\eqref{def:phi}).  The nonlinear wave's
amplitude, mean, wavenumber, and frequency are determined by the $r_j$
according to
\begin{equation}
\label{eq:3}
\begin{split}
a &= 2(r_2 - r_1)\,, \quad \overline{u} = r_1 + r_2 - r_3 +
2(r_3-r_1)\frac{E(m)}{K(m)}\,,\\
k &= \frac{\pi \sqrt{r_3 - r_1}}{\sqrt{6}
K(m)}\,, \quad \omega = \frac{k}{3}(r_1 + r_2 + r_3) \,.
\end{split}
\end{equation}

\begin{figure}
\centering
\includegraphics{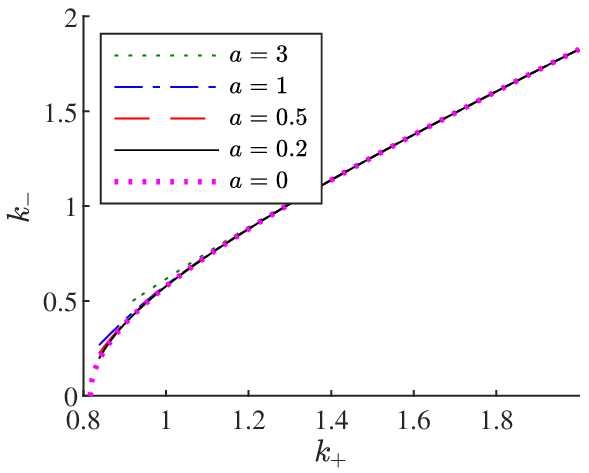}
\includegraphics{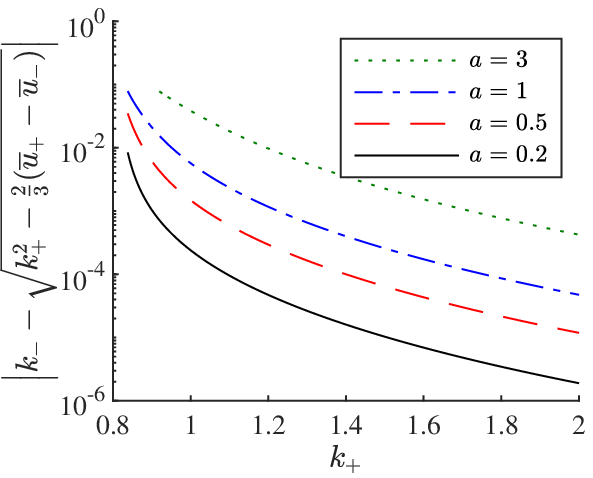}
\caption{Left plot: transmission relation between the left, $k_-$,
and right, $k_+$, wavenumbers for variable nonlinear plane wave
amplitude $a$.  Right plot: difference between the finite amplitude
wavepacket-mean flow transmission relation $k_-(k_+,a)$ and the zero
amplitude relation in Eq.~\eqref{sol:k}.  In both plots, the mean
flow satisfies $\overline{u}_- = 0$, $\overline{u}_+ = 1$. Even for
large amplitudes, the zero amplitude transmission relation is
accurate.}
\label{fig:transmission_a}
\end{figure}

First, we recover the already obtained transmission condition
\eqref{sol:k} by considering the Riemann problem \eqref{eq:t0} with
$r_2 \to r_1$ so that $a_\pm \to 0$ and we are in the linear wave
regime.  We also have $\ubar \to r_3$ and
$k \to 2\sqrt{r_3 - r_1}/\sqrt{6}$.  Solving for $r_1$, we obtain
\begin{equation}
\label{eq:5}
r_1 = \ubar - \frac{3}{2}k^2 \,,
\end{equation}
which is precisely the Riemann invariant $q$ in Eq.~\eqref{eq:qkdv}.
We also find $V_3 \to r_3$ so that the Riemann invariant $r_3 = \ubar$
in Eq.~\eqref{eq:1} ($j = 3$) satisfies the mean flow equation
\eqref{eq:ubar3} with $V(\ubar) = \ubar$ and admits the self-similar
solution $r_3 = x/t$.  The other two Riemann invariants coincide and
are constant $r_1 = r_2 = q$, so that evaluating \eqref{eq:5} on the
left and right states of the Riemann problem \eqref{eq:t0} yield the
transmission condition \eqref{sol:k}.  The plane wave-RW solution
involves variation only in the third characteristic field
(Eq.~\eqref{eq:1} with $j = 3$), which is an example of what is termed
a 3-wave in hyperbolic systems theory (see, e.g.,
\cite{el_dispersive_2016}).

\begin{figure}
\centering
\includegraphics{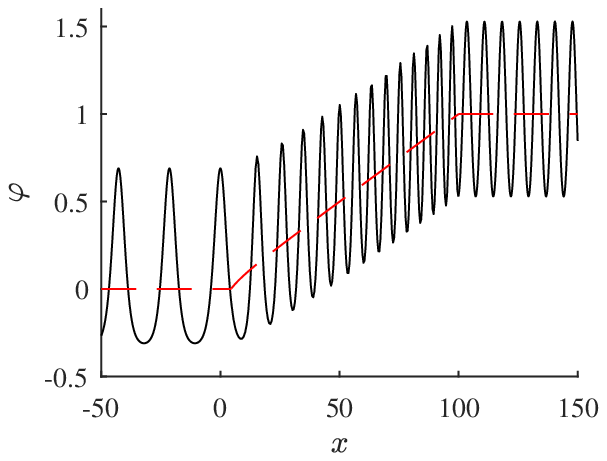}
\includegraphics{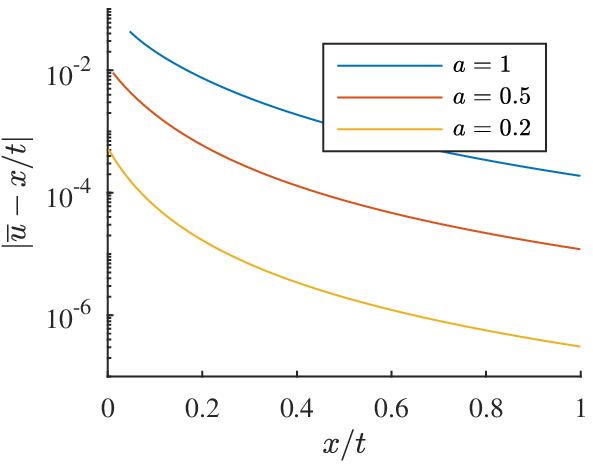}
\caption{Left plot: asymptotic solution for nonlinear plane
wave-RW interaction with $k_+ = 0.85$, $\ubar_- = 0$,
$\ubar_+ = 1$, and $a = 1$ so that $k_- \approx 0.294$ at
$t = 100$ (solid) and its corresponding mean flow variation
(dashed).  Right plot: difference between the self-similar RW
mean flow variation for $k_+ = 1$ in the 3-wave modulation
solution and the zero amplitude result $\ubar = x/t$. Even for
large amplitudes, the zero amplitude mean flow variation is
accurate.}
\label{fig:ubar_a}
\end{figure}

We now generalise this result to arbitrary, finite amplitude $a$ by
again considering the Riemann problem \eqref{eq:t0}.  The 3-wave
solution consists of constant Riemann invariants $r_1$ and $r_2$ with
$r_1 < r_2$.  Consequently, the nonlinear plane wave amplitude
$a = 2(r_2-r_1)$, no matter how large, is constant across the
self-similar dynamics of the mean flow because it is independent of
$r_3$.  First, we consider a right-incident nonlinear plane wave in
which the amplitude $a$, the right wavenumber $k_+$, and the left and
right mean flows $\ubar_- < \ubar_+$ are given---nonlinear plane
wave-RW interaction.  Then, the aim is to determine the left
wavenumber when a 3-wave solution exists.  These four constraints,
along with the equations in \eqref{eq:3}, determine $r_1$, $r_2$, and
$r_3^{\mp}$.  Then $k_-$ is not a free parameter and is determined in
terms of the four constraints, which we obtain by numerical solution
of the algebraic equations.  The absolute difference of the
relationship between $k_-$ and $k_+$ for variable amplitude $a$ and
the zero amplitude transmission condition \eqref{sol:k} is shown in
Fig.~\ref{fig:transmission_a}, left.  The error is surprisingly small,
even for very large plane wave amplitudes.  The self-similar mean flow
variation is determined by solving for $r_3(x/t)$ when
$V_- \le x/t \le V_+$ where $V_3(r_1,r_2,r_3) = x/t$ such that
$V_\pm = V_3(r_1,r_2,r_3^{\pm})$.  We perform this calculation
numerically and plot the difference between the mean flow variation
computed from the zero amplitude result, $\ubar = x/t$, and the 3-wave
solution's mean flow variation with nonzero amplitude in
Fig.~\ref{fig:transmission_a}, right.  The influence of the plane
wave's amplitude on the mean flow (i.e. the induced mean flow) is
almost negligible, even for very large amplitudes.

The process of obtaining $k_-$ in terms of the other flow parameters
($\ubar_\pm$, $k_+$, and $a$) does not require $\ubar_- < \ubar_+$,
therefore the calculation of the transmission relation for nonlinear
plane wave-DSW interaction is the same as for nonlinear plane wave-RW
interaction and we obtain the same transmission relation.  This is yet
another explanation of hydrodynamic reciprocity.  Note, however, that
the mean flow variation will necessarily be different, involving the
2-phase interaction of a DSW with a nonlinear plane wave.  We do not
study this problem here.

The conservation of the nonlinear plane wave's amplitude for
interactions with RW or DSW mean flows and the accuracy of the zero
amplitude transmission relation prediction for nonzero plane wave
amplitudes helps explain the numerically observed robustness of our
more general small amplitude wavepacket analysis described in the next
section.

\section{Interaction of linear wavepackets with unsteady hydrodynamic states}
\label{sec:numerics}

\subsection{Partial Riemann problem}
\label{sec:partial}

Having considered the model case of PW interaction with dispersive
hydrodynamic states, we now proceed with a more physically relevant
example of a similar interaction involving localised linear
wavepackets instead of PWs.  To model such an interaction, the Riemann
problem~\eqref{eq:mod}, \eqref{eq:t0} must be modified to take into
account the localised nature of the wavepacket.  To this end, we
introduce the {\it partial Riemann problem}
\begin{equation}
\label{eq:dis2}
(\ubar,k) =
\begin{cases}
(\ubar_-,k_-) &\text{if }x<0 \\
(\ubar_+,k_+) &\text{if }x>0
\end{cases}\,,\quad
a(x,0) = a_0 f(x-X_0) \,,
\end{equation}
where the amplitude profile is localised and centred at $x = X_0$.  We
take a Gaussian $f(y) = e^{-y^2/L_0^2}$ with width $L_0$. In what
follows, the position of the wavepacket, defined as a group velocity
line, is denoted by $X(t)$, so that according to~\eqref{eq:dis2},
$X(0)=X_0$.  While the wavepacket is localised, we consider a
sufficiently broad initial amplitude distribution that does not vary
significantly over one period $2\pi/k$ of the oscillation of the
carrier wave, $L_0 \gg 2\pi/k_\pm$, such that the amplitude modulation
is well described by~\eqref{eq:A}, see conditions~\eqref{eq:axa}.  We
also require $|X_0| \gg L_0$ so that the initial wavepacket is
well-separated from the initial step in the mean flow at the origin.

The quantities $k_+$ and $k_-$ here denote the dominant wavenumbers of
the wavepackets for $x>0$ and $x<0$ respectively. Note that, although
the wavepacket dominant wavenumber is defined only along the group
velocity line, we treat it here as a spatiotemporal field $k(x,t)$,
and the formulation~\eqref{eq:dis2} assumes the simultaneous presence
of two wavepackets at $t=0$ with dominant wavenumbers $k_-$ and $k_+$.
Still only one of them---we shall call it the incident wavepacket---is
physically realised due to the localised nature of the amplitude
distribution. The additional, fictitious wavepacket yields all the
transmission, trapping information of the incident wavepacket. See
\cite{maiden_solitonic_2018} for a similar extension made to define a
soliton amplitude field in the context of the soliton-mean flow
interaction problem.

The partial Riemann problem~\eqref{eq:mod}, \eqref{eq:dis2} implies
two possible interaction scenarios: (i) a right-incident interaction
where a wavepacket, initially placed at $X_0 > 0$, propagates with
group velocity $v_g^+=\ubar_+ - 3 k_+^2$ and enters either an
expanding hydrodynamic structure whose leading edge velocity is
$V(\ubar_+) > v_g^+$ (see~\eqref{edges_RW}, \eqref{edges_DSW}); (ii) a
left-incident interaction, where the wavepacket is initially placed at
$X_0<0$ so that the interaction only occurs if
$V(\ubar_-)< v_g^-=\ubar_- - 3k_-^2 $. It follows
from~\eqref{edges_RW}, \eqref{edges_DSW} that this can happen only for
a DSW but not for a RW.

The subsystem~\eqref{eq:ubar3}, \eqref{eq:k2} and~\eqref{eq:dis2} for
$\ubar$ and $k$ has already been solved in the previous section. The
simple wave solution of this problem is given by Eqs.~\eqref{sol:ubar}
and~\eqref{sol:q}, and thus the relation between $k_-$ and $k_+$
\eqref{sol:k} obtained for PWs, holds for the wavepacket-mean flow
interaction.  As a consequence, the wavepacket is subject to the
transmission condition~\eqref{sol:trans}.  The possible interaction
configurations are summarised in Table~\ref{tab:cond}.  The partial
Riemann problem \eqref{eq:kdv}, \eqref{eq:dis2} was solved numerically
to verify the relation~\eqref{sol:k} in Fig.~\ref{fig:k}. Its
numerical implementation is detailed in Appendix~\ref{app:num}.

\begin{table}
\centering
\begin{tabular}{p{2cm}p{5.3cm}p{5.3cm}}
Hydro. state &  Wavepacket $X_0 < 0$ &
Wavepacket $X_0> 0$ \\
\hline
RW & $\bullet$ no interaction & $\bullet$ transmitted if:
$k_+^2>2/3(\ubar_+-\ubar_-)$ $\bullet$ trapped in the RW otherwise
\\
\hline
DSW &$\bullet$ no interaction
if: $k_-^2>2/3(\ubar_--\ubar_+)$
$\bullet$ trapped in the DSW otherwise
&$\bullet$ always transmitted\\
\hline
\end{tabular}
\caption{\sl Configuration classification for
wavepacket---hydrodynamic state interaction.}
\label{tab:cond}
\end{table}

\subsection{Conservation of the integral of wave action}
\label{sec:cons_action}

We now proceed with the determination of the wavepacket amplitude
variation resulting from the interaction with dispersive hydrodynamic
states.  It is well known that KdV dispersion leads to wavepacket
broadening so that the amplitude $a(x,t)$ decreases during propagation
on a constant mean flow $\ubar_-$ or $\ubar_+$ in order to conserve
the integral $\int^\infty_{-\infty} a^2 dx$, which leads to the
standard dispersive decay estimate $a \sim t^{-1/2}$ for $t \gg 1$
\citep{whitham_linear_1999}.  Thus, we cannot expect the amplitude
transmission relation~\eqref{sol:a} derived for PWs to remain valid
for localised wavepackets.  To address this issue, instead of
considering the amplitude of the wave, we consider the integral of
wave action
\begin{equation}
\label{def:E}
E(t) = \int_{X_1(t)}^{X_2(t)} \frac{a(x,t)^2}{k(x,t)} d x \,,
\end{equation}
between two group lines
$dX_{1,2}/dt = v_g(k(x,t),\ubar(x, t)) \big|_{x=X_{1,2}(t)}$.  It then
follows from~\eqref{eq:lin_cons} and \eqref{eq:omega} that the
integral~\eqref{def:E} is conserved during linear wavepacket
propagation through a hydrodynamic state with varying mean flow
$\ubar (x,t)$ \citep{whitham_general_1965,bretherton_wavetrains_1968}.

\begin{figure}
\centering
\includegraphics{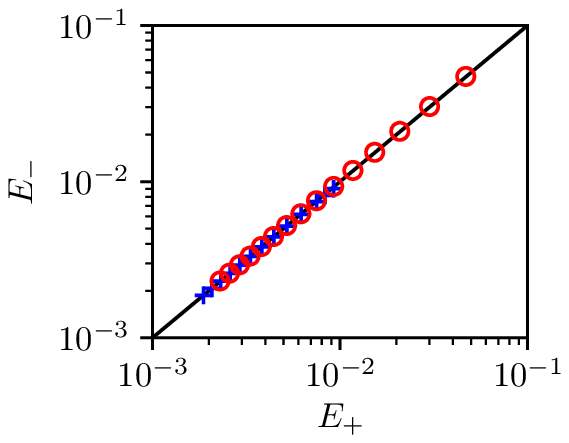}
\includegraphics{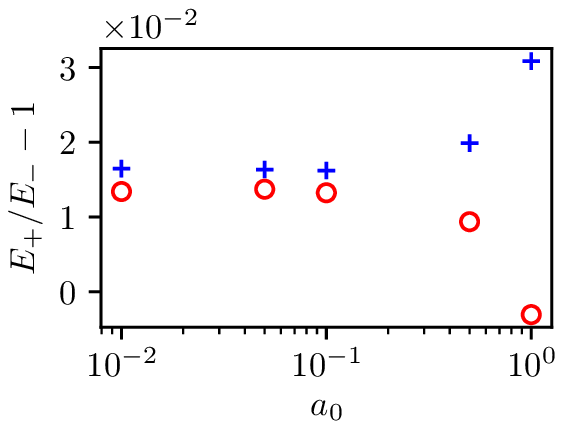}
\caption{\sl Left plot: comparison between the integral of wave action
computed before ($t=t_+$), and after ($t=t_-$), the transmission of
the wavepacket:
$E_\pm = (\int_{-\infty}^{+\infty} a(x,t_\pm) ^2 d x )/k_\pm $ for
the Riemann problem depicted in Fig.~\ref{fig:k}. The solid line
(\BlackSolid) corresponds to~\eqref{sol:inta2} and the crosses
(\BluePlus) or circles (\RedCircle) are obtained numerically from
linear wavepacket-RW or -DSW interaction, respectively with
$a_0 = 0.01$ and variable $k_+$. Right plot: relative error
$E_-/E_+-1$ for $k_+=0.88$ and different amplitudes $a_0$ of the
initial wavepacket.
}
\label{fig:inta2}
\end{figure}

If the incident wavepacket is transmitted and remains localised, we
can evaluate the integral~\eqref{def:E} before $(t < t_+)$ and after
($t > t_-$) the interaction when the wave is localised at the right or
the left of the hydrodynamic state, respectively, and where the
wavenumber field is uniform $k(x,t)=k_+$ for $t < t_+$ or
$k(x,t) = k_-$ for $t > t_-$ where $a(x,t) \neq 0$. Thus, the
conservation of the integral of wave action $E(t_-)=E(t_+)$ yields
\begin{equation}
\label{sol:inta2}
\frac{1}{k_-} \int_{-\infty}^{+\infty} a(x,t_-)^2 d x =
\frac{1}{k_+} \int_{-\infty}^{+\infty} a(x,t_+)^2 d x \,,
\end{equation}
where we replace the limits of integration $X_1$ and $X_2$ by
$-\infty$ and $+\infty$ since the wave\-packet is localised in space.
The relation~\eqref{sol:inta2} is valid in both linear wavepacket-RW
and -DSW interactions, as illustrated in
Fig.~\ref{fig:inta2}. Similar to the relation between $k_-$ and
$k_+$~\eqref{sol:k}, Eq.~\eqref{sol:inta2} holds beyond the small
amplitude limit of the wavepacket as displayed by the right plot in
Fig.~\ref{fig:inta2}.

In the case of a broad wavepacket of almost constant amplitude, we
have the following approximation:
\begin{equation}
\label{approx_wp}
\int_{-\infty}^{+\infty} a(x,t_\pm)^2 d x \approx a_\pm^2 L_\pm \,,
\end{equation}
where $a_\pm$ and $L_\pm$ are, respectively, the constant amplitude
and width of the wavepacket before and after interaction with a
hydrodynamic state. It follows from the wave conservation law that the
widths $L_\pm$ of the wavepackets on both sides of the hydrodynamic
state satisfy $L_-/k_- = L_+/k_+$ (see Eq.~\eqref{sol:width}) so that
Eqs.~\eqref{sol:inta2} and~\eqref{approx_wp} yield the approximate
conservation of amplitude:
$a_- \approx a_+$,
which agrees with Eq.~\eqref{sol:a} obtained for the limiting case of
PW-mean flow interaction.

\subsection{Wavepacket-rarefaction wave interaction}
\label{sec:wp-rw}

In this section, we consider in detail the interaction between a
wavepacket and a RW; as we already mentioned, the linear wavepacket
interacts with the RW only if initially $x=X_0 > 0$ (see
Tab.~\ref{tab:cond}). The fields $\ubar$ and $k$ are the solution of
the Riemann problem studied in Sec.~\ref{sec:adiabatic_rw}.  The
variation of $\ubar(x,t)$ is described by the
relation~\eqref{sol:ubar} with $V(\ubar)=\ubar$, and the variation of
$k(x,t)$ is given by:
\begin{equation}
k(x,t) = \sqrt{k_+^2 - 2/3(\ubar_+-\ubar(x,t))} \,,
\end{equation}
obtained through the conservation of the adiabatic
invariant~\eqref{eq:qkdv}.  The identification of a dominant
wavenumber when the wavepacket propagates inside the hydrodynamic
state implies that $k(x,t)$, or similarly $\ubar(x,t)$, is almost
constant across the wavepacket.  This latter condition is readily
satisfied for sufficiently large $t$ as the RW mean flow satisfies
$\ubar_x = 1/t$.

\begin{figure}
\centering
\includegraphics{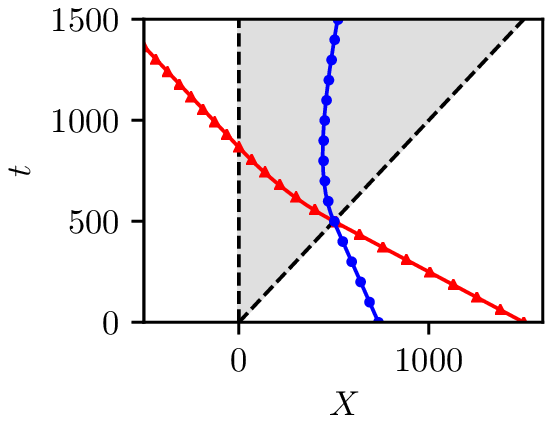}
\includegraphics{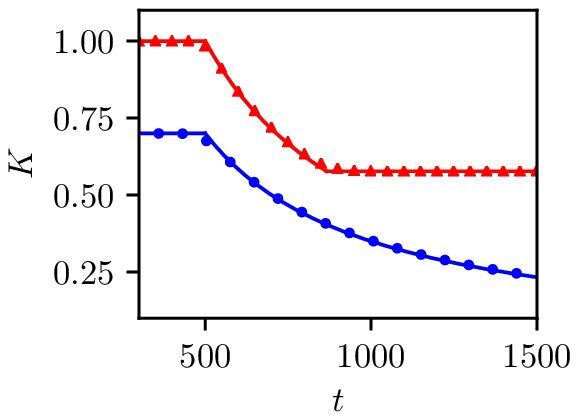}
\caption{\sl Left plot: trajectories of wavepacket-RW
interaction. Solid lines correspond to the solution~\eqref{sol:X} and
markers to the numerical trajectory. The triangles (\RedTriangle)
correspond to the transmission configuration for $k_+=1$ and the dots
(\BlueDisc) correspond to the trapping configuration when $k_+=0.7$.
The RW edges $x=\ubar_\pm t$ are represented by dotted lines
(\BlackDot).  Right plot: the corresponding temporal variation of the
wavepacket wavenumber. Solid lines correspond to the
solution~\eqref{def:K} and markers to the numerical result.}
\label{fig:traj_rw}
\end{figure}

\begin{figure}
\centering
\includegraphics{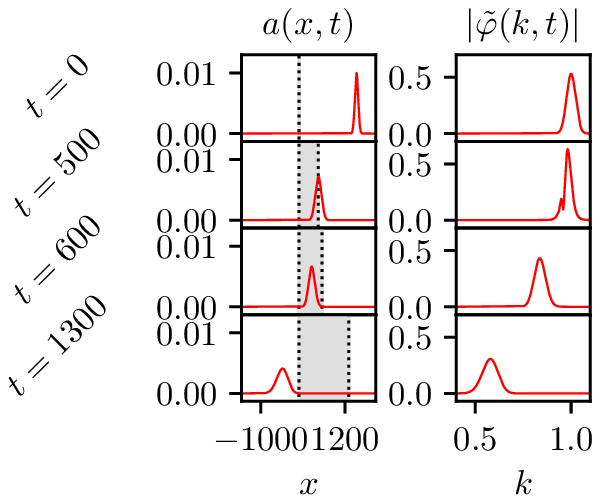}
\includegraphics{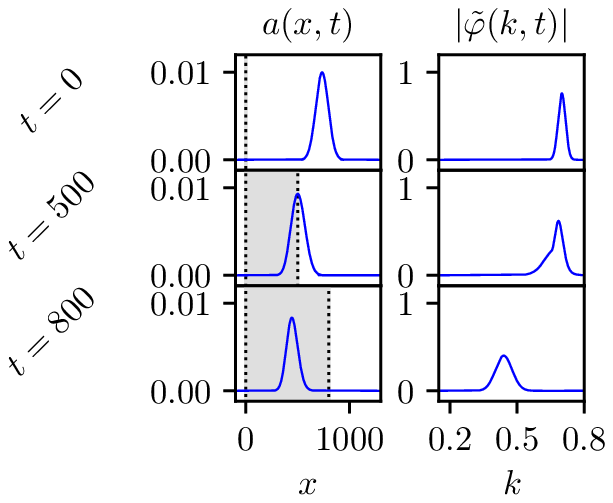}
\caption{\sl Numerical evolution of wavepacket-RW interaction in the
case of transmission (left plot) and trapping (right plot). The
first column displays the extracted wavepacket envelope
$a(x,t)$. The positions of the RW edges are shown by dotted lines
(\BlackDot).  The second column displays the amplitude of the
wavepacket's Fourier transform, denoted $\tilde \varphi$.}
\label{fig:snap_RW}
\end{figure}

Since the wavepacket propagates with the group velocity
$v_g(k,\ubar) = \ubar - 3k^2$, its position $X(t)$ satisfies the
characteristic equation
\begin{equation}
\label{eq:X}
\frac{d X}{d t} = v_g\left(k(X,t),\ubar(X,t) \right) \,,\quad X(0) =
X_0 > 0 \,.
\end{equation}
The integration of~\eqref{eq:X} yields
\begin{equation}
\label{sol:X}
X(t)=\begin{cases}
v_g(k_+,\ubar_+)\, t +X_0& \text{for } 0\le t \le t_+\\
(\ubar_+-\tfrac32 k_+^2)t +X_0 t_+/(2 t) & \text{for } t_+\le t \le t_-\\
v_g(k_-,\ubar_-)\, t+3 k_-^2t_- & \text{for } t_-\le t
\end{cases}\,,
\end{equation}
where $t_+=X_0/(3 k_+^2)$ and $t_-=X_0/(3 k_+ k_-)$. Hence, during the
interaction with the RW, the temporal variation of the dominant
wavepacket wavenumber $K(t)$ along the group velocity line is given by
\begin{equation}
\label{def:K} K(t) = k(X(t),t) = k_+ t_+/t \,.
\end{equation}
The wavepacket trajectory described by Eq.~\eqref{sol:X} is compared
with the numerically observed trajectory in Fig.~\ref{fig:traj_rw} for
two different configurations: transmission ($k_+>k_c$,
cf.~\eqref{sol:trans}) and trapping ($k_+<k_c$).  Snapshots of the
envelope $a(x,t)$ of the wavepacket field $\varphi(x,t)$ and the
absolute value of its Fourier transform $|\tilde \varphi(k,t)|$ are
presented in Fig.~\ref{fig:snap_RW}.  The numerical procedure
implemented to extract $\varphi(x,t)$ from the full numerical solution
of~\eqref{eq:kdv} is explained in Appendix~\ref{app:num}.

In Fig.~\ref{fig:snap_RW}, the wavepacket shape in Fourier space
slightly deviates from Gaussian when it enters the leading RW edge at
$t=500$.  However, the wavepacket recovers its Gaussian form when it
is fully inside the RW and after exiting the RW.

While the wavepacket propagates at constant velocity over a
nonmodulated mean flow $\ubar(x,t)=\ubar_+$ or $\ubar(x,t)=\ubar_-$,
the wavepacket decelerates during the propagation inside the RW for
$t_+ < t < t_-$.  Note that acceleration/deceleration here is
understood as the increasing/decreasing of the group speed
$|v_g(X,t)|$. If the transmission condition~\eqref{sol:trans} is not
satisfied, $\lim_{t \to \infty} K(t) = 0$, and the incident wavepacket
gets trapped inside the RW as its velocity $v_g=\ubar-3K^2$ converges
asymptotically to the local background velocity $\ubar$.  Moreover,
the wavepacket amplitude decays indefinitely, following the
conservation of wave action~\eqref{def:E}, and the wavepacket
eventually gets absorbed by the RW (see Fig.~\ref{fig:snap_RW},
right).

We now draw certain parallels between the trapping of linear waves in
RWs and the effect of so-called wave blocking in counter-propagating,
inhomogeneous steady currents $U(x)<0$, where the wavenumber $k(x)$
also varies following the inhomogeneities of the current (recall the
discussion in Sec.~\ref{sec:intro}). In this case, the adiabatic
variation of the wavenumber is simply described by the conservation of
the frequency $\omega(k(x);U(x))={\rm const}$.  In contrast to wave
trapping due to wavepacket-RW interaction considered here, wave
blocking in the counter-propagating current is accompanied by a
decrease in wave\-packet wavelength and an increase in amplitude, until
it reaches the stopping velocity $v_g=0$ at some finite wavenumber.

\begin{figure}
\centering
\includegraphics{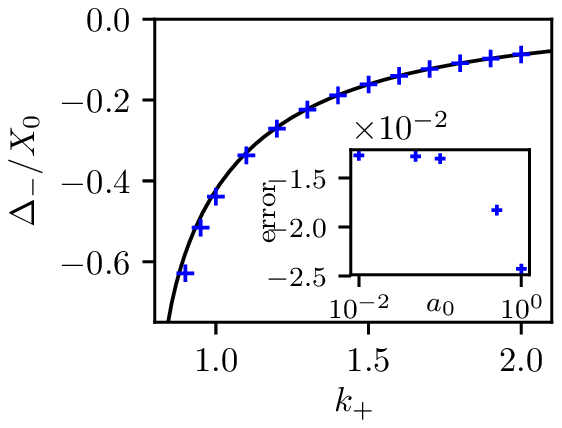}
\includegraphics{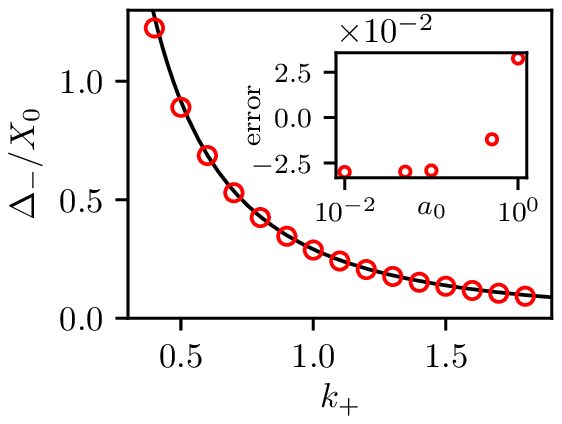}
\caption{\sl Numerical determination of the normalised phase shift
$\Delta_-/X_0$ for wavepacket-RW interaction (left plot) and
wavepacket-DSW interaction (right plot). In the two plots, the
markers (pluses \BluePlus~for the RW interaction and circles
\RedCircle~for the DSW interaction) correspond to the numerical
simulation and the solid line to the analytical
prediction~\eqref{sol:Delta2}. The inset plots correspond to the
relative error $\Delta_-^{\rm num.}/\Delta_- - 1$ between
$\Delta_-^{\rm num.}$ determined numerically and $\Delta_-$ given by
the relation~\eqref{sol:Delta2}. $\Delta_-^{\rm num.}$ is obtained
for different initial wavepacket amplitudes $a_0$, $k_+=1.2$ for
wavepacket-RW interaction (\BluePlus) and $k_+=0.88$ for
wavepacket-DSW interaction (\RedCircle).}
\label{fig:Delta}
\end{figure}

The trajectory of the wavepacket displayed in Fig.~\ref{fig:traj_rw}
shows that the wavepacket undergoes refraction due to its interaction
with the RW. In the transmission configuration, this results in both a
speed shift and a phase shift of the transmitted wavepacket.  The
phase of the wave after its transmission is equal to
$X_-= 3k_-^2t_- \neq X_0$ (cf. Eq.~\eqref{sol:X} for $t>t_-$), so that
the phase shift $\Delta_- = X_- - X_0$ is
\begin{equation}
\label{sol:Delta2}
\frac{\Delta_-}{X_0} = \frac{k_-}{k_+} - 1\,.
\end{equation}
This result can also be obtained from the second adiabatic invariant
$pA$ in Eq.~\eqref{sol:pA} where $A= \delta k$ as in
Eq.~\eqref{sol:deltak}. Viewing the wavepacket as part of a fictitious
periodic train of wavepackets, we recognise that the relative position
of the wavepackets post ($x = X_-$) and pre ($x = X_0$) interaction is
inverse to the relative beating wavenumber shift
$X_-/X_0=\delta k_+/\delta k_- = k_-/k_+$.  Since $k_-<k_+$, the phase
shift is negative in the considered situation. The
formula~\eqref{sol:Delta2} is precisely~\eqref{sol:Delta} when we
identify $L_+$ with $X_0$, using the second adiabatic invariant
$p \delta k$. Fig.~\ref{fig:Delta} displays the phase shift computed
numerically for different wavenumbers $k_+$, which agrees with
relation~\eqref{sol:Delta2}. In addition, the
relation~\eqref{sol:Delta2} holds for large amplitude wavepackets,
just as the relations~\eqref{sol:k} and~\eqref{sol:inta2} do.

\subsection{Wavepacket-DSW interaction}
\label{sec:wp-dsw}

We now consider the more complex case of wavepacket-DSW
interaction. Such an interaction is generally described by two-phase
KdV modulation theory, which is quite technical, with modulation
equations given in terms of hyperelliptic integrals
\citep{flaschka_multiphase_1980}. The mean field approach adopted here
enables us to circumvent these technicalities by employing the
approximate modulation system~\eqref{eq:mod} that yields simple and
transparent analytic results that, as we will show, agree extremely
well with direct numerical simulations. More broadly, the notion of
hydrodynamic reciprocity described in Sec.~\ref{sec:adiabatic_dsw} can
be utilised without approximation to make specific predictions for
wavepacket-DSW interaction for $t>0$ based on wavepacket-RW
interaction for $t<0$.

As already mentioned, wavepacket-DSW interaction admits two basic
configurations (see Table \ref{tab:cond}): the transmission
configuration, arising when $X_0>0$ and applicable to any incident
wavenumber $k_+>0$, and the trapping configuration, when $X_0<0$ and
the incident wavenumber $k_->0$ is sufficiently small.  The variation
of $\ubar(x,t)$ inside the DSW is given by $\ubar=V^{-1}(x/t)$ (see
Eq.~\eqref{sol:ubar}) where the characteristic velocity $V(\ubar)$ is
defined by~\eqref{eq:V(u)}.

\begin{figure}
\centering
\includegraphics{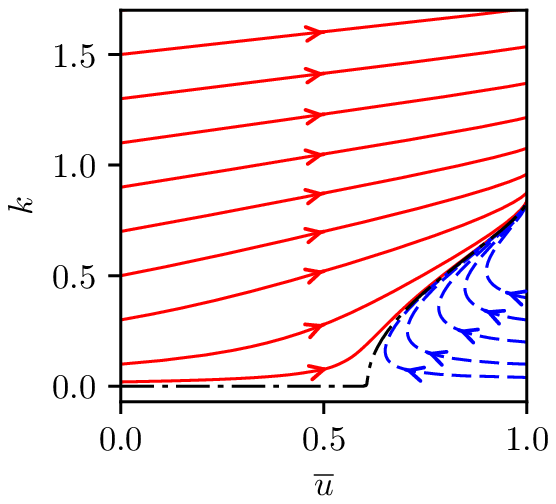}
\includegraphics{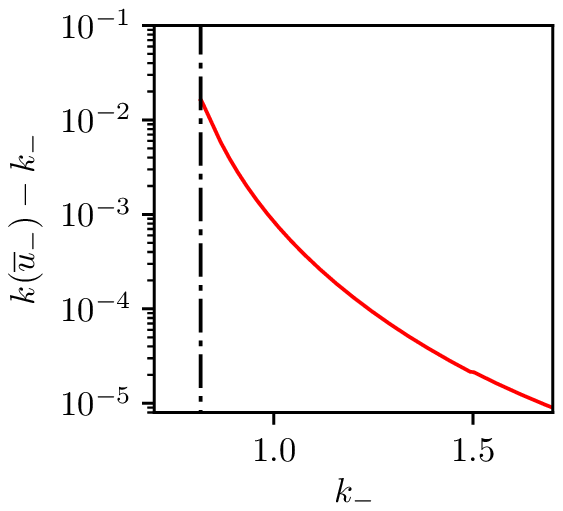}
\caption{\sl Left plot: wave curves for wavepacket-DSW interaction
obtained by numerical integration of Eq.~\eqref{eq:ku} with
$(\ubar_-,\ubar_+)=(1,0)$.  Solid curves (\RedSolid) correspond to
transmission configurations ($k(\ubar_-)> \sqrt{2/3}$), dashed
curves (\BlueDash) to trapped configurations
($k(\ubar_-)< \sqrt{2/3}$) and the dash-dotted curve (\BlackDashDot)
to the limiting case $k(\ubar_-) \simeq \sqrt{2/3}$.  The arrows
correspond to the direction associated with propagation of the
wavepacket. Right plot: deviation of the predicted transmitted
wavenumber $k(\overline{u}_-)$ from the actual value $k_-$ obtained
from Eq.~\eqref{sol:k} and hydrodynamic reciprocity, as a function
of $k_-$ with $\overline{u}_- = 1$, $\overline{u}_+ = 0$.  The
vertical dash-dotted line is the minimum transmitted wavenumber
$k_-=\sqrt{2/3}$.}
\label{fig:ku}
\end{figure}

\begin{figure}
\centering
\includegraphics{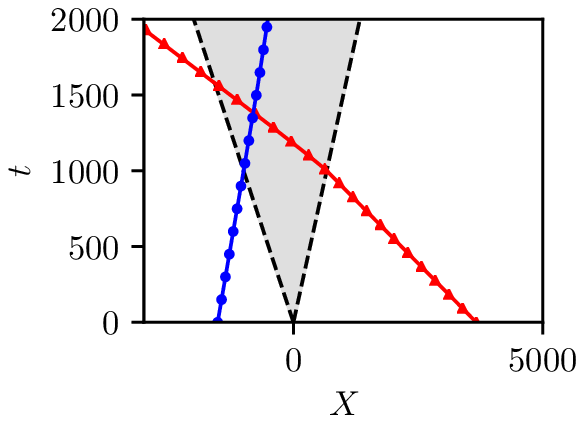}
\includegraphics{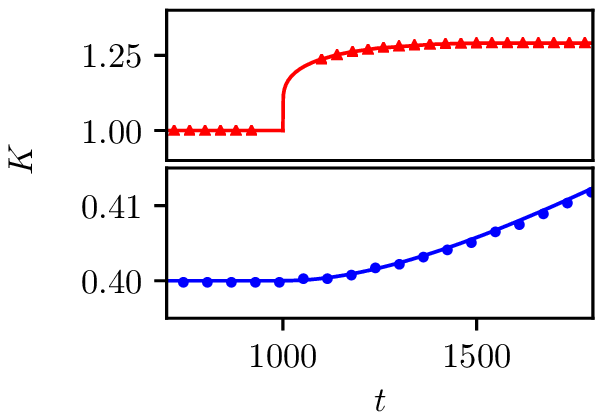}
\caption{\sl Left plot: trajectories of wavepacket-DSW
interaction. Solid lines correspond to semi-analytical solutions
obtained by solving~\eqref{eq:X}, \eqref{eq:ku} and markers to the
numerical resolution of the corresponding Riemann problem.  The
 triangles (\RedTriangle) correspond to
the transmission configuration (left propagating wavepacket with
$k_+=1$) and the dots (\BlueDisc)
correspond to the trapping configuration (right propagating
wavepackets with $k_-=0.4$).  The DSW edge trajectories $x=s_- t$
and $x=s_+ t$ are displayed as dotted lines (\BlackDot): in both
cases, we set $\ubar_-=1$ and $\ubar_+=0$. Right plot: corresponding
temporal variation of the wavenumbers along the wavepacket
trajectories. Solid lines correspond to the semi-analytical solution
$K(t) = k(\ubar(X(t),t))$ where $k(\ubar)$ has been determined by
solving~\eqref{eq:ku} numerically.}
\label{fig:traj_dsw}
\end{figure}

The variation of the wavepacket's wavenumber field $k(x,t)$ is given
by the adiabatic invariant $q$, which can be obtained by integrating
the differential form $\Xi$ in Eq.~\eqref{diff_form}.  This
differential form vanishes on the group velocity characteristic
$dx /dt = v_{g}$, yielding a relation between $k$ and $\ubar$
specified by the ODE
\begin{equation}
\label{eq:ku}
\frac{d k}{d \ubar} =
\frac{\omega_{\ubar}(k,\ubar)}{V(\ubar)-\omega_k(k,\ubar)} \,,
\end{equation}
with the boundary condition $k(\ubar_+)=k_+$ if $X_0>0$ or
$k(\ubar_-)=k_-$ if $X_0<0$.  Note that equation~\eqref{eq:ku} for
$V(\ubar)=\ubar$ arises in the DSW fitting method where it determines
the locus of the KdV DSW harmonic edge, see \cite{el_resolution_2005,
el_dispersive_2016}.  Here, it has a different meaning and does not
appear to be amenable to analytical solution because of the presence
of elliptic integrals in the function $V(\ubar)$. We therefore
solve~\eqref{eq:ku} numerically.  Once the relation $k(\ubar)$ has
been determined, the $(x,t)$-dependence of the wavenumber inside the
DSW is $k = k(\ubar(x,t)) = k(V^{-1}(x/t))$. The wavepacket trajectory
in the $(x,t)$-plane is obtained by solving~\eqref{eq:X} with the
already determined $\ubar(x,t)$ and $k(x,t)$. The results of our
semi-analytical computations are presented in Figs.~\ref{fig:ku} and
\ref{fig:traj_dsw}.

Fig.~\ref{fig:ku} (left plot) displays the wave curves $k(\ubar)$
obtained from the numerical integration of~\eqref{eq:ku}, that can be
interpreted as wavepacket trajectories in the parameter space
$(\ubar,k)$.  The evolution of the wavepacket's wavenumber
$K(t)=k(\ubar(X/t))$ along a wave curve is then described by an ODE
\begin{equation}
\label{eq:dK}
\frac{d K}{d t} = \frac{-K}{V'(\ubar) t} \,,
\end{equation}
obtained by combining $\ubar
=V^{-1}(X/t)$, Eqs.~\eqref{eq:X} and~\eqref{eq:ku}. Since the
characteristic speed
$V(\ubar)$~\eqref{eq:ubar} of the Gurevich-Pitaevskii modulation
equation is a decreasing function of
$\ubar$ (cf.~Fig.~\ref{fig:uz}), Eq.~\eqref{eq:dK} shows that the
wavepacket's wavenumber is increasing during its propagation inside
the DSW, in contrast to wavepacket-RW interaction for which
$V'(\ubar)=1>0$.

We now verify that the obtained integral curves for wavepacket-DSW
interaction are consistent with the transmission relation~\eqref{sol:k}
for PW-RW interaction, as required by hydrodynamic reciprocity. In the
transmission configuration, where
$k(\ubar_-)>k_c = \sqrt{\frac23 (\ubar_--\ubar_+)}$ (see
\eqref{sol:trans}), the wave curve $k(\ubar)$ is represented by a
solid curve in Fig.~\ref{fig:ku}, left that connects $\ubar=\ubar_+$
to $\ubar=\ubar_-$ and, for a given incident wavenumber $k_+$, the
transmitted wavenumber is obtained by evaluating $k(\ubar_-)$. Figure
\ref{fig:ku}, right shows the comparison of the transmitted wavenumber
$k(\ubar_-)$ evaluated by the above semi-analytical procedure with the
value $k_-=\sqrt{k_+^2+ \frac23 (\ubar_--\ubar_+)}$ obtained from the
wavepacket-RW transmission condition Eq.~\eqref{sol:k} by invoking
hydrodynamic reciprocity.  The agreement confirms the validity of the
mean field approximation and its consistency with hydrodynamic
reciprocity.

As expected, the behaviour of wave curves $k(\ubar)$ is drastically
different for the trapping configuration, when $k(\ubar_-)<k_c$.  In
this case, the curves $k(\ubar)$ (represented by dashed curves in
Fig.~\ref{fig:ku}) do not connect $\ubar=\ubar_-$ to $\ubar=\ubar_+$
anymore, implying that the wavepacket initially placed at $x=X_0<0$
cannot reach the mean flow $\ubar = \ubar_+$ (trapping).
Interestingly, these trapping wave curves $k(\ubar)$ are multi-valued,
which implies that wavepackets with initial parameters
$\ubar =\ubar_-, k<k_c$ will return, asymptotically as $t \to \infty$,
to the DSW harmonic edge where $\ubar =\ubar_-$. More specifically,
the point $\ubar=\ubar_-,k=k_c$ plays the role of an attractor in the
parameter space for trapping configurations, such that all the trapped
wavepackets' wavenumbers converge to the same value $k_c$ with
time. This filtering behaviour is unusual and drastically different
from wavepacket-RW trapping (see Sec.~\ref{sec:wp-rw}), where the
wavepacket trajectory is
single-valued and  $k \to 0$ as $ t \to \infty$.\\

\begin{figure}
\centering
\includegraphics{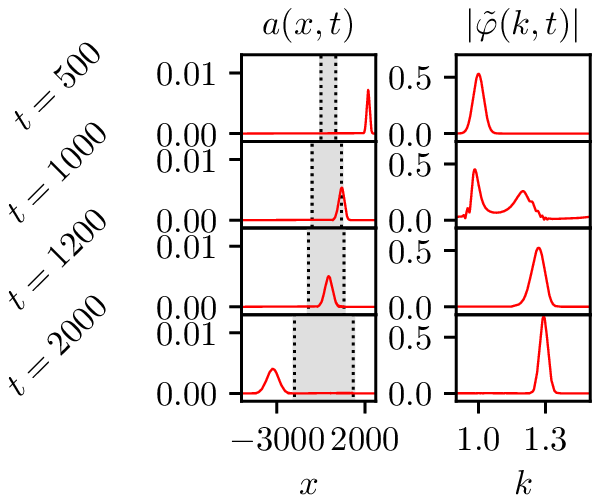}
\hspace{-0.5cm}
\includegraphics{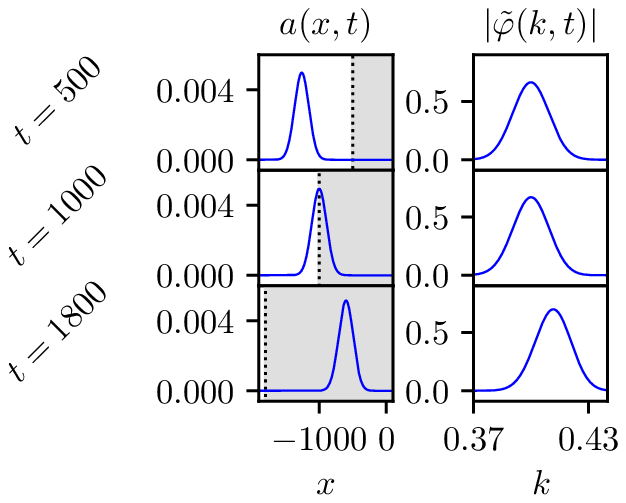}
\caption{\sl Numerical evolution of wavepacket-DSW interaction, in the
transmitted configuration (left plot) and in the trapped
configuration (right plot).  The first column displays the
wavepacket's envelope amplitude $a(x,t)$ and the positions of the RW
edges are indicated by dotted lines (\BlackDot).  The second column
displays the amplitude of the wavepacket's Fourier transform.}
\label{fig:snap_DSW}
\end{figure}

We now compare the wavepacket dynamics obtained through our modulation
analysis with the numerical solution of the KdV equation with initial
conditions given by the partial Riemann data~\eqref{eq:dis2} (see
Appendix~\ref{app:num} for details of the numerical procedure employed
to trace the dynamics of a wavepacket inside a
DSW). Fig.~\ref{fig:traj_dsw} displays trajectories for the
transmitted and trapped wavepacket configurations, and snapshots of
the envelope $a(x,t)$ and the Fourier transform of $\varphi(x,t)$ for
the corresponding numerical simulation are presented in
Fig.~\ref{fig:snap_DSW}. The agreement of the numerical simulations
with the analytical predictions in Fig.~\ref{fig:traj_dsw} represents
a further confirmation of the mean field approximation employed in the
derivation of the basic ODE~\eqref{eq:ku}.

Similar to the interaction with a RW, the group velocity of the linear
wavepacket is not constant inside the DSW---but now the wavepacket
accelerates in the transmission case and simultaneously experiences a
wavenumber increase. Here, however, the determination of the
wavenumber $K(t)$ is not everywhere possible in the numerical
simulation, see Fig.~\ref{fig:traj_dsw}. This becomes obvious when one
follows the evolution of the amplitude of the Fourier transform
$|\tilde \varphi(k,t)|$ of the linear field $\varphi(x,t)$ along with
the envelope $a(x,t)$ of the field itself
(cf. Fig.~\ref{fig:snap_DSW}).  Initially in our simulations, both
distributions have a Gaussian shape, but the Fourier transform of the
amplitude distribution loses its unimodality when the wavepacket
initially interacts with the leading, soliton, edge of the DSW (see
Fig.~\ref{fig:snap_DSW}, left plot at $t=1000$).  In fact, close to
the soliton edge the mean flow gradient $\ubar_x$ is logarithmically
singular \citep{gurevich_nonstationary_1974, el_resolution_2005}, see
Fig.~\ref{fig:uz}, and the wavenumber field $k(x,t)$ varies
significantly over the extent of the wavepacket.  As a result, we are
no longer in a position to define a nearly monochromatic carrier wave
in the wavepacket.  Figure~\ref{fig:snap_DSW} shows that the
quasi-monochromatic wavepacket structure is recovered when the
interaction with the DSW edge is over.  Its wavenumber is still
described by the adiabatic analytical result while the wavepacket
propagates in the region where $\ubar$ is almost constant over the
wavepacket extension. Ultimately $K=k_-$, where $k_-$ is given by the
relation~\eqref{sol:k}, when the wavepacket is no longer interacting
with the DSW due to hydrodynamic reciprocity.

The above described logarithmic divergence of the mean field gradient
is absent in wavepacket-RW interactions considered in the previous
section, where $\ubar= x/t$ inside the hydrodynamic state such that
$k_x \propto \ubar_x$ remains finite but exhibits a discontinuity. We
still observe a similar, slight deviation from monochromaticity in
wavepacket-RW interaction at the initial stage,
cf.~Fig.~\ref{fig:snap_RW} $t=500$, with $\ubar$ varying significantly
along the extension of the wavepacket. This behaviour, expectedly,
does not appear in the wavepacket-DSW trapping interaction (see
Fig.~\ref{fig:snap_DSW}, right panel), where the wavepacket, coming
from the left of the hydrodynamic state, only interacts with the
slowly varying part of the mean flow.

Similar to wavepacket-RW interaction, we consider the phase shift of
the wavepacket transmitted through the DSW. The numerical results are
presented in Fig.~\ref{fig:Delta} (right plot, lower panel) and are in
agreement with the value predicted analytically in
Eq.~\eqref{sol:Delta2} via hydrodynamic reciprocity.

We note in conclusion that the trapping configuration is somewhat more
difficult to treat analytically using the mean field approach employed
in other cases. Although the trajectory, as well as the dominant
wavenumber of the wavepacket, can be approximately described by our
theory for short time evolution (see Fig.~\ref{fig:traj_dsw}), we
numerically observe that the dynamics of the DSW are no longer
decoupled from the dynamics of the linear wave, and the distinction
between the two structures becomes less and less pronounced after a
sufficiently long time.

\section{Conclusions and Outlook}
\label{sec:conclusions}

In the context of shallow water theory, we have introduced a general
mathematical framework in which to study the interaction of linear
wavepackets with unsteady nonlinear dispersive hydrodynamic states:
rarefaction waves (RWs) and dispersive shock waves (DSWs) or undular
bores.  We use a combination of classical Whitham modulation theory
and the mean field approximation to derive a new, extended modulation
system that describes the dispersive dynamics of a linear wavepacket
coupled to the nonlinear, long-wave dynamics of the mean flow in the
hydrodynamic state. The mean field equation coincides with the long
wave limit of the original dispersive equation when the hydrodynamic
state is slowly varying (RW) but has a more complicated structure for
rapidly oscillating states (DSWs).  We show that the extended
modulation system admits a convenient, general diagonalisation
procedure that reveals conserved adiabatic invariants during
wavepacket evolution through a slowly evolving mean flow. These
adiabatic invariants predict transmission relations and trapping
conditions for the incident wavepacket. They also imply the
hydrodynamic reciprocity property whereby wavepacket interactions with
RWs and DSWs exhibit the same transmission/trapping conditions. This
enables the circumvention of the complicated analysis of DSW mean
field behaviour in order to take advantage of the available
wavepacket-RW relations to describe the transmission through a DSW or
predict wavepacket trapping inside a DSW.  This study has been
performed using the KdV equation as a prototypical example, although
the integrability properties of the KdV equation were not invoked. The
developed theory can be extended to other models supporting
multi-scale nonlinear dispersive wave propagation.

While the modulation equations~\eqref{eq:mod} are formally valid only
in the limit of vanishingly small amplitude waves,
$\varphi \ll \max(\ubar) - \min(\ubar)$, the numerical simulations
demonstrated that the resulting transmission relation \eqref{sol:k}
between $k_+$ and $k_-$ also holds for waves of moderate amplitudes
$a \sim |\ubar_+-\ubar_-|$.  This becomes important for establishing
the applicability of the transmission relation~\eqref{sol:k} to the
actual water wave system modelled by the KdV equation for long surface
waves in shallow water. In the context of the full water wave model,
the KdV equation describes the propagation of weakly nonlinear
perturbations to the water surface, so the field $u(x,t)$ already
constitutes a small quantity compared to the total depth. Then,
considering linearised waves within the KdV approximation would trim
the small amplitude limit even further. The robustness of the linear
modulation theory results for larger amplitude waves ensures here that
the relation~\eqref{sol:k} remains valid for realistic physical
situations where the incident and transmitted waves do not necessarily
have small amplitudes {\it within the KdV approximation}, but are of
the same order $O(\ubar_+-\ubar_-)$ as the nonlinear hydrodynamic
states described by KdV.

The developed modulation theory of linear wave-mean flow interactions
could be applied to the interaction of wind-generated short waves with
shallow water undular bores in coastal ocean environments.  This
scenario could readily be tested in wave tank experiments
\citep{hammack_korteweg-vries_1974, treske_undular_1994,
frazao_undular_2002, trillo_observation_2016, rousseaux_novel_2016}
where slowly varying mean flows and undular bores have been generated.
Another promising application area is to the interaction of small
amplitude, short waves with rising and ebbing tide generated mean
flows in internal ocean waves.  For example, the observed physical
parameters pertaining to large-scale undular bores and the
Brunt-V\"ais\"al\"a linear dispersion relation in
\citep{scotti_shoaling_2008} conform to the assumptions underlying the
analysis presented in this paper.  The KdV equation can only describe
weakly nonlinear internal waves \citep{helfrich_long_2006}.
Nevertheless, the theory developed here can readily be generalised to
models that capture strongly nonlinear phenomena occurring in a
variety of coastal areas
\citep{scotti_shoaling_2008,harris_intermittent_2017,li_seasonal_2018}.

This theory can also be utilised in many physical contexts beyond
classical fluid mechanics. In particular, similar to the soliton-mean
flow interaction theory very recently developed in
\cite{maiden_solitonic_2018}, it can be applied to a broad range of
dispersive hydrodynamic systems describing wave propagation in
nonlinear optics and condensed matter physics, opening perspectives
for experimental observation of the various interaction scenarios
studied here.  In fact, the linear wavepacket transmission and
trapping configurations can be interpreted as hydrodynamic wavepacket
scattering, a dispersive wave counterpart of hydrodynamic soliton
tunnelling
\citep{maiden_solitonic_2018,sprenger_hydrodynamic_2018}. In both
cases, the role of a barrier or a scatterer is played by a
large-scale, evolving hydrodynamic state that satisfies the same
equation as the soliton (wavepacket).  Finally, we mention the
actively developing field of analogue gravity (see
\cite{barcelo_analogue_2011} and references therein), where the
effects of dispersive wave trapping studied here may find interesting
interpretations.

A major challenge for the modern theory of dispersive hydrodynamics is
to develop a stability theory for dispersive shock waves.  The
linearisation about a DSW involves a differential operator with both
spatially and temporally varying coefficients, presenting significant
challenges for its further analysis.  The work presented here suggests
that perturbations involving sufficiently short wavelengths can be
successfully described via wave-mean flow interaction, thus greatly
simplifying the stability analysis in this regime.

The developed theory admits generalisations and opens interesting
perspectives. It can be extended to physically relevant systems of
``KdV type'', such as the asymptotically equivalent, long wave
Benjamin-Bona-Mahony equation, the Gardner equation for internal
waves, the Kawahara equation for capillary-gravity waves or the
viscous fluid conduit equation.  Extensions to systems with a
nonconvex hyperbolic flux or nonconvex linear dispersion relation may
prove fruitful because nonconvexity is known to lead to profound
effects in dispersive hydrodynamics: undercompressive and contact DSWs
\cite{el_dispersive_2017}, expansion shocks \citep{el_expansion_2016},
DSW implosion \citep{lowman_dispersive_2013} and the existence of
resonant and travelling DSWs \citep{sprenger_shock_2017}.  Another
natural extension of this work is to the study of linear wave-mean
flow interaction in the framework of integrable and non-integrable
bidirectional systems such as the defocusing nonlinear Schr\"odinger
equation, the Serre system for fully nonlinear shallow water waves
\citep{serre_contribution_1953}, and the Choi-Camassa system for fully
nonlinear internal waves \cite{choi_fully_1999}.

The abstract, basic modulation system~\eqref{eq:lin_cons} has been
extensively used in the theory of phase modulations that reveal
dispersive deformations arising near coalescing characteristics (see
\cite{bridges_symmetry_2017, ratliff_whitham_2016} and references
therein). At present, this theory does not include variations of the
mean flow.  The modulation system that couples modulations of the
wavepacket to mean field variations studied here could also be useful
for further development of phase modulation theory.

Finally we mention one more area where an appropriate extension of the
developed modulation theory could prove useful. It is related to the
fundamental problem of mean flow-turbulence interaction (see,
e.g., \cite{Falkovich_2016} and references therein).  A possible
connection between weak limits of nonlinear dispersive waves and
turbulence theories was conjectured by \cite{Lax_1991}.  Extensions of
the modulation theory approach described here to multi-dimensional
linear and weakly nonlinear waves provides a plausible entry into this
connection.

\section*{Acknowledgements}

The work of TC and GAE was partially supported by the EPSRC grant
EP/R00515X/1. The work of MAH was partially supported by NSF grants
CAREER DMS-1255422 and DMS-1816934. Authors gratefully acknowledge
valuable discussions with Nicolas Pavloff, Roger Grimshaw and Sylvie
Benzoni-Gavage. Authors also thank Laboratoire de Physique Th\'eorique
et Mod\`eles Statistiques (Universit\'e Paris-Saclay) where this work
was initiated.

\appendix

\section{Wavepacket-DSW interaction: numerical resolution}
\label{app:num}

The initial step~\eqref{eq:dis2} of the partial Riemann problem is
implemented numerically by the function:
\begin{equation}
\label{def:riem1}
u(x,t=0) = \ubar_0(x) + \varphi_0(x)\,,
\end{equation}
with:
\begin{equation}
\label{def:u0}
\begin{split}
&\ubar_0(x) = \frac{\ubar_+-\ubar_-}{2} \tanh \left( \frac{x}{\xi} \right) +
\frac{\ubar_++\ubar_-}{2} \,,\\
&\varphi_0(x) = a_0 \exp \left( -\frac{(x-X_0)^2}{L_0^2} \right)
\cos \left[ k_\pm(x-X_0) \right] \,,
\end{split}
\end{equation}
where we set $\xi=5$, $L_0 = 120/k_\pm$ and, except where otherwise
stated, $a_0=0.01$. The values for $a_0$ and $L_0$ are chosen such
that $\varphi_0$ has small amplitude and is a sufficiently broad
wavepacket.  The problem \eqref{eq:kdv}, \eqref{def:riem1},
\eqref{def:u0} is then solved numerically with homogeneous Neumann
boundary conditions. The numerical scheme adopted here to solve the
KdV equation is explicit, where the space derivatives are approximated
using centered finite differences and the time integration is
performed with the 4th order Runge-Kutta method.  Note that to solve
\eqref{eq:kdv}, \eqref{def:riem1}, \eqref{def:u0} with
$(\ubar_-,\ubar_+)=(1,0)$ for $t<0$ in Fig.~\ref{fig:k}, we solve the
equivalent problem with $(\ubar_-,\ubar_+)=(0,1)$ for $t>0$.

In order to determine the variations of the wavepacket $\varphi(x,t)$,
we also numerically solve the Riemann problem with the initial
condition $u(x,t=0)= \ubar_0(x)$ such that we obtain for $t>0$,
$u(x,t) = u_{\rm H.S.}(x,t)$. Thus, supposing that the numerical
solution $u(x,t)$ of~\eqref{eq:kdv}, \eqref{def:riem1}, \eqref{def:u0}
can be put in the form~\eqref{sol:gen}, we obtain the variations of
the wavepacket $\varphi(x,t)$ by evaluating the difference:
\begin{equation}
\varphi(x,t) = u(x,t) - u_{\rm H.S.}(x,t) \,.
\end{equation}
We then extract from the wavepacket amplitude $a(x,t)$, the position
of the wavepacket
\begin{equation}
\label{def:xnum}
X(t) = \frac{\int_{-\infty}^{+\infty}
\,a^2(x,t) \,x\,  d x}{\int_{-\infty}^{+\infty}a^2(x,t) \, d x }\,,
\end{equation}
and from the spatial Fourier transform
$\tilde \varphi(k,t) = {\cal F}[\varphi(x,t)]$, the wavepacket
dominant wavenumber
\begin{equation}
\label{def:knum}
K(t) = \frac{\int_{-\infty}^{+\infty}
|\tilde \varphi(k,t)|^2  \,k\, d k}{\int_{-\infty}^{+\infty}  |\tilde
\varphi(k,t)|^2 \, d k} \,.
\end{equation}
Note that here, the position~\eqref{def:xnum} and the dominant
wavenumber~\eqref{def:knum} correspond to average quantities instead
of the pointwise maxima of $a(k,t)$ and $\tilde \varphi(k,t)$,
respectively, which are not uniquely defined in some situations.  When
the wavepacket is Gaussian, the quantities~\eqref{def:xnum},
\eqref{def:knum} are equivalent to the conventional definitions of the
wavepacket position and dominant
wavenumber.

\begin{figure}
\centering
\includegraphics{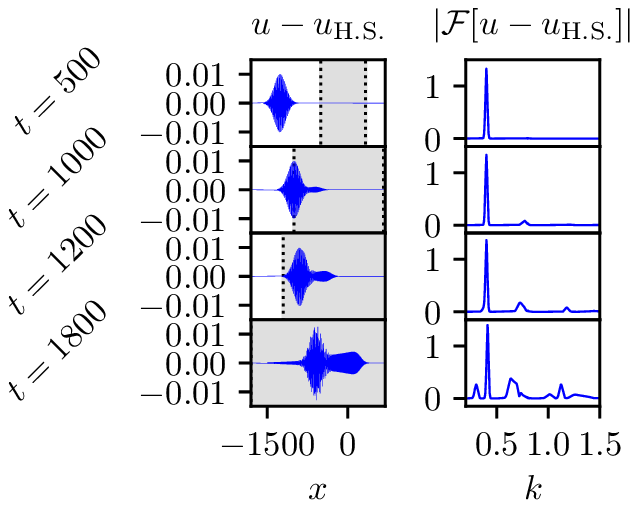}
\includegraphics{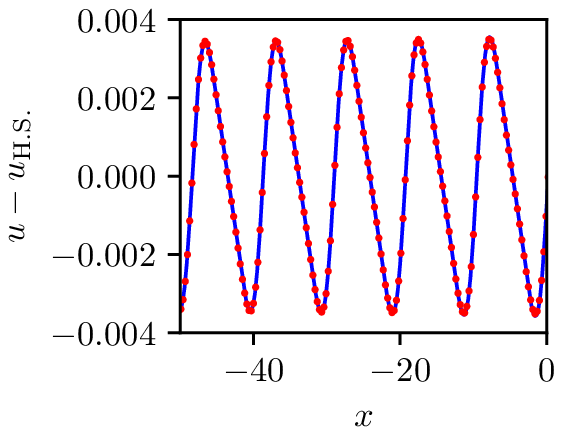}
\caption{\sl Left plot: numerical evolution of the field
$u-u_{\rm H.S.}$ and its spatial Fourier transform wavepacket in
the trapping interaction with a DSW.  Right plot: the line
(\BlueSolid) corresponds to a zoom in of the oscillation emerging
to the right of the wavepacket at $t=1800$. The dots (\RedDisc)
correspond to the derivative $\partial_x u_{\rm H.S.}(x,t)$,
rescaled in order to compare with $u-u_{\rm H.S.}$.}
\label{fig:snap_DSW_trap}
\end{figure}

The ansatz~\eqref{sol:gen} proves to be inadequate to describe the
variations of $u(x,t)$ in the trapping interaction with a DSW.  We
observe that the field $u(x,t) - u_{\rm H.S.}(x,t)$ no longer
corresponds to a quasi-monochromatic wavepacket, and exhibits
additional small harmonic excitations as in
Fig.~\ref{fig:snap_DSW_trap}.
We identify this deviation from unimodality as a local phase shift of
the DSW.  Indeed, it is known that a soliton interacting with a
dispersive wavetrain is phase shifted with respect to its free
propagation \citep{ablowitz_note_1982}, and a similar phenomenon could
happen for DSWs which are approximately rank-ordered soliton
trains. Thus, a schematic solution of the Riemann problem should read:
\begin{equation}
u(x,t) = u_{\rm H.S.}(x-\delta(x,t),t) + \varphi(x,t) \,,
\end{equation}
where $\delta(x,t) \ll 1$ corresponds to the small phase-shift induced
by the wavepacket-DSW interaction. Yet we determine $\varphi(x,t)$
numerically by computing the difference $u(x,t) - u_{\rm H.S.}(x,t)$
such that
\begin{equation}
\label{sol:phi_num2}
u(x,t) - u_{\rm H.S.}(x,t) \approx \varphi(x,t)  - \frac{\partial u_{\rm
H.S.}(x,t)}{\partial x}\delta(x,t) \,.
\end{equation}
The shape of the oscillations of this new linear structure seems to
correspond qualitatively to $\partial_x u_{\rm H.S.}(x,t)$, cf.
Fig.~\ref{fig:snap_DSW_trap}. 

The evolution of the wavepacket $\varphi$ is recovered numerically by
eliminating the term corresponding to the DSW phase shift
in~\eqref{sol:phi_num2}. This phase-shift contribution can be clearly
identified at an early stage of the evolution ($t \sim 10^{-3}$) as
the {\it non-adiabatic} generation of harmonic excitations in the
spatial Fourier transform of $u - u_{\rm H.S.}$.  The filtered signal
of Fig.~\ref{fig:snap_DSW_trap} is displayed in
Fig.~\ref{fig:snap_DSW}. It is surprising that, even if the DSW
dynamics are slightly perturbed by the wavepacket's propagation, the
wavepacket dynamics are well described by the theory developed here,
which confirms, once again, the mean field assumption for
wavepacket-DSW interaction.


\begin{thebibliography}{46}
\expandafter\ifx\csname natexlab\endcsname\relax\def\natexlab#1{#1}\fi

\bibitem[Ablowitz \& Benney(1970)]{ablowitz_evolution_1970} {\sc
Ablowitz, M.~J. \& Benney, D.~J.} 1970 The evolution of multi-phase
modes for nonlinear dispersive waves. {\em Stud. Appl. Math.\/} {\bf
49}, 225--238.

\bibitem[Ablowitz \& Kodama(1982)]{ablowitz_note_1982} {\sc Ablowitz,
M.~J. \& Kodama, Y.} 1982 Note on asymptotic solutions of the
{Korteweg}-de {Vries} equation with solitons. {\em
Stud. Appl. Math.\/} {\bf 66}, 159--170.

\bibitem[Abramowitz \& Stegun(1972)]{abramowitz_handbook_1972} {\sc
Abramowitz, M. \& Stegun, I.~A.} 1972 {\em Handbook of {mathematical}
{functions} with {formulas}, {graphs}, and {mathematical}
{tables}\/}. New York: Dover Publications.

\bibitem[Barber(1949)]{barber_behaviour_1949} {\sc Barber, N.~F.} 1949
The behaviour of waves on tidal streams. {\em Proc. R. Soc. Lon. Ser. A\/} {\bf 198}~(1052), 81--93.

\bibitem[Barcel\'o {\em et~al.\/}(2011)]{barcelo_analogue_2011} {\sc
Barcel\'o, C., Liberati, S., \& Visser, M. } 2011 Analogue gravity
{\em Living Rev. Relativ.} {\bf 14}: 3.

\bibitem[{Bona} {\em et~al.\/}(2002){Bona}, {Chen} \&
{Saut}]{bona_boussinesq_2002} {\sc {Bona}, {Chen} \& {Saut}} 2002
Boussinesq {equations} and {other} {systems} for {small}-{amplitude}
{long} {waves} in {nonlinear} {dispersive} {media}. {i}: {derivation}
and {linear} {theory}. {\em J. Nonlinear Sci.\/} {\bf 12}~(4),
283--318.

\bibitem[Bona {\em et~al.\/}(2004)Bona, Chen \&
Saut]{bona_boussinesq_2004} {\sc Bona, J.~L., Chen, M. \& Saut, J.-C.}
2004 Boussinesq equations and other systems for small-amplitude long
waves in nonlinear dispersive media: {II}.  {The} nonlinear
theory. {\em Nonlinearity\/} {\bf 17}~(3), 925--952.

\bibitem[Bretherton(1968)]{bretherton_propagation_1968} {\sc
Bretherton, F.~P.} 1968 Propagation in slowly varying waveguides. {\em
Proc. R. Soc. Lon. Ser. A\/} {\bf 302}~(1471), 555--576.

\bibitem[Bretherton \& Garrett(1968)]{bretherton_wavetrains_1968} {\sc
Bretherton, F.~P. \& Garrett, C.~J.~R.} 1968
Wavetrains in inhomogeneous moving media. {\em
Proc. R. Soc. Lon. Ser. A\/} {\bf 302}~(1471), 529--554.

\bibitem[Brevik \& Aas(1980)]{brevik_flume_1980} {\sc Brevik, I.
\& Aas, B.} 1980 Flume experiment on waves and currents.
{I}. {Rippled} bed. {\em Coastal Engineering\/} {\bf 3}, 149--177.

\bibitem[Bridges(2017)]{bridges_symmetry_2017} {\sc Bridges, T.~J.}
2017 {\em Symmetry, {phase} {modulation} and {nonlinear}
{waves}\/}. Cambridge: Cambridge University Press.

\bibitem[B\"uhler(2009)]{buhler_waves_2009} {\sc B\"uhler, O.} 2009
{\em Waves and mean flows\/}. Cambridge: Cambridge University Press.

\bibitem[Choi \& Camassa(1999)]{choi_fully_1999} {\sc Choi, W. \&
Camassa, R.} 1999 Fully nonlinear internal waves in a two-fluid
system. {\em J. Fluid Mech.\/} {\bf 396}, 1--36.

\bibitem[Dobrokhotov \& Maslov(1981)]{dobrokhotov_finite-zone_1981}
{\sc Dobrokhotov, S.~Y. \& Maslov, V.~P.} 1981 Finite-zone,
almost-periodic solutions in {WKB} approximations. {\em
J. Math. Sci.\/} {\bf 16}~(6), 1433--1487.

\bibitem[El(2005)]{el_resolution_2005} {\sc El, G.~A.} 2005 Resolution
of a shock in hyperbolic systems modified by weak dispersion. {\em
Chaos\/} {\bf 15}, 037103.

\bibitem[El \& Hoefer(2016)]{el_dispersive_2016} {\sc El, G.~A. \&
Hoefer, M.~A.} 2016 Dispersive shock waves and modulation theory. {\em
Physica D\/} {\bf 333}, 11--65.

\bibitem[El {\em et~al.\/}(2016)El, Hoefer \&
Shearer]{el_expansion_2016} {\sc El, G.~A., Hoefer, M.~A. \& Shearer,
M.} 2016 Expansion shock waves in regularized shallow-water
theory. {\em Proc. R. Soc. Lon. Ser. A\/} {\bf 472}~(2189), 20160141.

\bibitem[El {\em et~al.\/}(2017)El, Hoefer \&
Shearer]{el_dispersive_2017} {\sc El, G.~A., Hoefer, M.~A. \& Shearer,
M.} 2017 Dispersive and diffusive-dispersive shock waves for nonconvex
conservation laws. {\em SIAM Rev.\/} {\bf 59}~(1), 3--61.

\bibitem[Evans(1955)]{evans_pneumatic_1955} {\sc Evans, J.~T.} 1955
Pneumatic and similar breakwaters. {\em Proc. R. Soc. Lon. Ser. A\/}
{\bf 231}~(1187), 457--466.

\bibitem[Falkovich(2016)]{Falkovich_2016} {\sc Falkovich, G.} 2016
Interaction between mean flow and turbulence in two dimensions. {\em
Proc. R. Soc. Lon. Ser. A\/} {\bf 472}~(2191), 20160287.

\bibitem[Flaschka {\em et~al.\/}(1980)Flaschka, Forest \&
McLaughlin]{flaschka_multiphase_1980} {\sc Flaschka, H., Forest,
M.~G. \& McLaughlin, D.~W.} 1980 Multiphase averaging and the inverse
spectral solution of the {Korteweg}-de {Vries} equation. {\em
Comm. Pure Appl. Math.\/} {\bf 33}, 739--784.

\bibitem[Frazao \& Zech(2002)]{frazao_undular_2002} {\sc Frazao,
S.~S. \& Zech, Y.} 2002 Undular bores and secondary
waves -Experiments and hybridfinite-volume modelling. {\em
J. Hydraul. Res.\/} {\bf 40}~(1), 33--43.

\bibitem[Gallet \& Young(2014)]{gallet_refraction_2014} {\sc Gallet,
B. \& Young, W.~R.} 2014 Refraction of swell by surface currents {\em
J. Mar. Res.\/} {\bf 72}, 105--126.

\bibitem[Garrett(1968)]{garrett_interaction_1968} {\sc Garrett,
C.~J.~R.} 1968 On the interaction between internal gravity waves and a
shear flow. {\em J. Fluid Mech.\/} {\bf 34}~(4), 711--720.

\bibitem[Grimshaw(1975)]{grimshaw_nonlinear_1975} {\sc Grimshaw, R.}
1975 Nonlinear internal gravity waves in a rotating fluid.  {\em
J. Fluid Mech.\/} {\bf 71}~(3), 497--512.

\bibitem[Grimshaw(1984)]{grimshaw_wave_1984} {\sc Grimshaw, R.} 1984
Wave {action} and {wave}-{mean} {flow} {interaction}, with
{application} to {stratified} {shear} {flows}. {\em Annu. Rev. Fluid
Mech.\/} {\bf 16}~(1), 11--44.

\bibitem[Gurevich \& Pitaevskii(1974)]{gurevich_nonstationary_1974}
{\sc Gurevich, A.~V. \& Pitaevskii, L.~P.} 1974 Nonstationary
structure of a collisionless shock wave. {\em Sov. Phys. JETP\/} {\bf
38}~(2), 291--297, translation from Russian of A.~V. Gurevich and
L.~P. Pitaevskii, Zh. Eksp.  Teor. Fiz. 65, 590-604 (August 1973).

\bibitem[Haller \& Tuba \"Ozkan-Haller(2007)]{haller_waves_2007} {\sc
Haller, M.~C. \& Tuba \"Ozkan-Haller, H.} 2007 Waves on unsteady
currents. {\em Phys. Fluids\/} {\bf 19}~(12), 126601.

\bibitem[Hammack \& Segur(1974)]{hammack_korteweg-vries_1974}
{\sc Hammack, J.~L. \& Segur, H.} 1974 The {Korteweg}-de {Vries}
equation and water waves. {Part} 2. {Comparison} with experiments. {\em
J. Fluid Mech.\/} {\bf 65}~(2), 289--314.

\bibitem[Hammack \& Segur(1978{\natexlab{{\em
a\/}}})]{hammack_korteweg-vries_1978} {\sc Hammack, J.~L. \& Segur,
H.} 1978{\natexlab{{\em a\/}}} The {Korteweg}-de {Vries} equation and
water waves. {Part} 3. {Oscillatory} waves. {\em J. Fluid Mech.\/}
{\bf 84}~(2), 337--358.

\bibitem[Hammack \& Segur(1978{\natexlab{{\em
b\/}}})]{hammack_modelling_1978} {\sc Hammack, J.~L. \& Segur, H.}
1978{\natexlab{{\em b\/}}} Modelling criteria for long water
waves. {\em J. Fluid Mech.\/} {\bf 84}~(2), 359--373.

\bibitem[Harris \& Decker(2017)]{harris_intermittent_2017} {\sc
Harris, J.~C. \& Decker, L.} 2017 Intermittent large amplitude
internal waves observed in {Port Susan, Puget Sound}. {\em
Estuar. Coast. Shelf S.\/} {\bf 194}, 143--149.

\bibitem[Hayes(1970)]{hayes_conservation_1970} {\sc Hayes, W.~D.}
1970 Conservation of action and modal wave action.  {\em
Proc. R. Soc. Lon. Ser. A\/} {\bf 320}~(1541), 187--208.

\bibitem[Helfrich \& Melville(2006)]{helfrich_long_2006} {\sc
Helfrich, K.~R. \& Melville, W.~K.} 2006 Long nonlinear internal
waves. {\em Ann. Rev. Fluid Mech.\/} {\bf 38}, 395--425.

\bibitem[Hughes(1978)]{hughes_effect_1978} {\sc Hughes, B.~A.} 1978
The effect of internal waves on surface wind waves 2.  {Theoretical}
analysis. {\em J. Geophys. Res. C\/} {\bf 83}~(C1), 455--465.

\bibitem[Irvine(1985)]{irvine_kinematics_1985} {\sc Irvine, D.~E.}
1985 The {kinematics} of {short} {wave} {modulation} by {long}
{waves}. In {\em The {ocean} {surface}: {wave} {breaking}, {turbulent}
{mixing} and {radio} {probing}\/} (ed. Y. Toba \& H.  Mitsuyasu),
pp. 129--134. Dordrecht: Springer Netherlands.

\bibitem[Johnson(1947)]{johnson_refraction_1947} {\sc Johnson, J.~W.}
1947 The refraction of surface waves by currents. {\em Eos,
Trans. Amer. Geophys. Union\/} {\bf 28}~(6), 867--874.

\bibitem[Kamchatnov(2000)]{kamchatnov_nonlinear_2000} {\sc Kamchatnov,
A.~M.} 2000 {\em Nonlinear periodic waves and their modulations: an
introductory course\/}. Singapore: World Scientific.

\bibitem[Lai {\em et~al.\/}(1989)Lai, Long \&
Huang]{lai_laboratory_1989} {\sc Lai, R.~J., Long, S.~R. \& Huang,
N.~E.} 1989 Laboratory studies of wave-current interaction:
{Kinematics} of the strong interaction.  {\em J. Geophys. Res. C\/}
{\bf 94}~(C11), 16201--16214.

\bibitem[Lannes(2013)]{lannes_water_2013} {\sc Lannes, D.} 2013 {\em
The water waves problem\/}. Providence, RI: American Mathematical
Society.

\bibitem[Lax(1991)]{Lax_1991} {\sc Lax, P.~D.}  1991 The zero
dispersion limit, a deterministic analogue of turbulence. {\em
Comm. Pure Appl. Math.\/} {\bf 44}~(8--9), 1047--1056.

\bibitem[Li, Pawlowicz, \& Wang(2018)]{li_seasonal_2018} {\sc Li, L.,
Pawlowicz, R. \& Wang, C.} 2018 Seasonal variability and
generation mechanisms of nonlinear internal waves in the strait of
{Georgia}. {\em J. Geophys. Res.-Oceans\/} {\bf 123}~(8),
5706--5726.

\bibitem[Longuet-Higgins \&
Stewart(1960)]{longuet-higgins_changes_1960} {\sc Longuet-Higgins,
M.~S. \& Stewart, R.~W.} 1960 Changes in the form of short gravity
waves on long waves and tidal currents. {\em J. Fluid Mech.\/} {\bf
8}~(4), 565--583.

\bibitem[Longuet-Higgins \&
Stewart(1961)]{longuet-higgins_changes_1961} {\sc Longuet-Higgins,
M.~S. \& Stewart, R.~W.} 1961 The changes in amplitude of short
gravity waves on steady non-uniform currents. {\em J. Fluid Mech.\/}
{\bf 10}~(4), 529--549.

\bibitem[Lowman \& Hoefer(2013)]{lowman_dispersive_2013} {\sc Lowman,
N.~K. \& Hoefer, M.~A.} 2013 Dispersive shock waves in viscously
deformable media. {\em J. Fluid Mech.\/} {\bf 718}, 524--557.

\bibitem[Luke(1966)]{luke_perturbation_1966} {\sc Luke, J.~C.} 1966 A
perturbation method for nonlinear dispersive wave problems. {\em
Proc. R. Soc. Lon. Ser. A\/} {\bf 292}~(1430), 403--412.

\bibitem[Maiden {\em et~al.\/}(2018)Maiden, Anderson, Franco, El \&
Hoefer]{maiden_solitonic_2018} {\sc Maiden, M.~D., Anderson, D.~V.,
Franco, N.~A., El, G.~A.  \& Hoefer, M.~A.} 2018 Solitonic
{dispersive} {hydrodynamics}: {theory} and {observation}. {\em
Phys. Rev. Lett.\/} {\bf 120}, 144101.

\bibitem[Mei {\em et~al.\/}(2005)Mei, Stiassnie \&
Yue]{mei_theory_2005} {\sc Mei, C.~C., Stiassnie, M. \& Yue, D.~K.-P.}
2005 {\em Theory and applications of ocean surface
waves\/}. Hackensack, N.J: World Scientific.

\bibitem[Peregrine(1976)]{peregrine_interaction_1976} {\sc Peregrine,
D.~H.} 1976 Interaction of {water} {waves} and {currents}. {\em
Adv. Appl. Mech.\/} {\bf 16}, 9--117.

\bibitem[Onorato {\em et~al.\/}(2011)Onorato, Proment \&
Toffoli]{onorato_triggering_2011} {\sc Onorato, M., Proment, D. \&
Toffoli, A.} 2011 Triggering rogue waves in opposing currents {\em
Phys. Rev. Lett.\/} {\bf 107}, 184502.

\bibitem[Peregrine \& Jonsson(1983)]{peregrine_interaction_1983} {\sc
Peregrine, D.~H. \& Jonsson, I.~G.} 1983 Interaction of waves and
currents.  {\em Tech. Rep.\/} 83-6. CERC, U.S. Army Corps of
Engineers.

\bibitem[Phillips(1980)]{phillips_dynamics_1980} {\sc Phillips, O.~M.}
1980 {\em The dynamics of the upper ocean\/}. Cambridge; New York:
Cambridge University Press.

\bibitem[Ratliff \& Bridges(2016)]{ratliff_whitham_2016} {\sc Ratliff,
D.~J. \& Bridges, T.~J.} 2016 Whitham modulation equations, coalescing
characteristics, and dispersive {Boussinesq} dynamics.  {\em Physica
D\/} {\bf 333}, 107--116.

\bibitem[Rousseaux {\em et~al.\/}(2008)Rousseaux, Mathis, Maissa,
Philbin \& Leonhardt]{rousseaux_observation_2008} {\sc Rousseaux, G.,
Mathis, C., Ma\"issa, P., Philbin, T.~G. \& Leonhardt, U.} 2008
Observation of negative-frequency waves in a water tank: a classical
analogue to the {Hawking} effect? {\em New J.  Phys.\/} {\bf 10}~(5),
053015.

\bibitem[Rousseaux {\em et~al.\/}(2010)Rousseaux, Maissa, Mathis,
Coullet, Philbin \& Leonhardt]{rousseaux_horizon_2010} {\sc Rousseaux,
G., Ma\"issa, P., Mathis, C., Coullet, P., Philbin, T.~G. \& Leonhardt,
U.} 2010 Horizon effects with surface waves on moving water. {\em New
J. Phys.\/} {\bf 12}~(9), 095018.

\bibitem[Rousseaux {\em et~al.\/}(2016)Rousseaux, Mougenot, Chatellier,
David \& Calluaud]{rousseaux_novel_2016} {\sc Rousseaux,
G., Mougenot, J.-M., Chatellier, L., David, L., \& Calluaud,
D.} 2016 A novel method to generate tidal-like bores in the
laboratory. {\em Euro. J. Mech. B\/} {\bf 55}, 31--38.

\bibitem[Schneider \& Wayne(2002)]{schneider_rigorous_2002} {\sc
Schneider, Guido \& Wayne, C.~Eugene} 2002 The {rigorous}
{approximation} of {long}-{wavelength} {capillary}-{gravity}
{waves}. {\em Arch. Rat.  Mech. Anal.\/} {\bf 162}~(3), 247--285.

\bibitem[Scotti, Beardsley, Butman \&
Pineda(2008)]{scotti_shoaling_2008} {\sc Scotti, A., Beardsley, R.~C.,
Butman, B \& Pineda, J.} 2008 Shoaling of nonlinear internal waves in
{Massachusetts Bay}. {\em J. Geophys. Res.\/} {\bf 113}~(C8), C08031.

\bibitem[Serre(1953)]{serre_contribution_1953} {\sc Serre, F} 1953
Contribution \`a l’ \'etude des \'ecoulements permanents et variables
dans les canaux. {\em La Houille Blanche\/} ~(3, 6), 374--388,
830--872.

\bibitem[Sprenger \& Hoefer(2017)]{sprenger_shock_2017} {\sc Sprenger,
P. \& Hoefer, M.~A.} 2017 Shock {waves} in {dispersive}
{hydrodynamics} with {nonconvex} {dispersion}. {\em SIAM
J. Appl. Math.\/} {\bf 77}, 26--50.

\bibitem[Sprenger {\em et~al.\/}(2018)Sprenger, Hoefer \&
El]{sprenger_hydrodynamic_2018} {\sc Sprenger, P., Hoefer, M.~A. \&
El, G.~A.} 2018 Hydrodynamic optical soliton tunneling. {\em Phys.
Rev. E\/} {\bf 97}~(3).

\bibitem[Taylor(1955)]{taylor_action_1955} {\sc Taylor, G.~I.} 1955
The action of a surface current used as a breakwater. {\em
Proc. R. Soc. Lon. Ser. A\/} {\bf 231}~(1187), 466--478.

\bibitem[Tolman(1990)]{tolman_influence_1990} {\sc Tolman, H.~L.} 1990
The {influence} of {unsteady} {depths} and {currents} of {tides} on
{wind}-{wave} {propagation} in {shelf} {seas}. {\em J. Phys.
Oceanogr.\/} {\bf 20}~(8), 1166--1174.

\bibitem[Treske(1994)]{treske_undular_1994} {\sc Treske, A.} 1994
Undular bores (favre-waves) in open channels -Experimental
studies. {\em J. Hydraul. Res.\/} {\bf 32}~(3), 355--370.

\bibitem[Trillo {\em et~al.\/}(2016)Trillo, Klein, Clauss \&
Onorato]{trillo_observation_2016} {\sc Trillo, S., Klein, M., Clauss,
G.~F. \& Onorato, M.} 2016 Observation of dispersive shock waves
developing from initial depressions in shallow water.  {\em Physica
D\/} {\bf 333}, 276--284.

\bibitem[Unna(1941)]{unna_white_1941} {\sc Unna, P.~J.~H.} 1941
“{White} horses”. {\em Nature\/} {\bf 148}~(3747), 226--227.

\bibitem[Whitham(1965{\natexlab{{\em a\/}}})]{whitham_general_1965}
{\sc Whitham, G.~B.} 1965{\natexlab{{\em a\/}}} A general approach to
linear and non-linear dispersive waves using a {Lagrangian}. {\em
J. Fluid Mech.\/} {\bf 22}~(02), 273--283.

\bibitem[Whitham(1965{\natexlab{{\em b\/}}})]{whitham_non-linear_1965}
{\sc Whitham, G.~B.} 1965{\natexlab{{\em b\/}}} Non-linear dispersive
waves.  {\em Proc. R. Soc. Lon. Ser. A\/} {\bf 283}, 238--261.

\bibitem[Whitham(1999)]{whitham_linear_1999} {\sc Whitham, G.~B.} 1999
{\em Linear and nonlinear waves\/}. New York, NY: Wiley.

\bibitem[{Whitham \& Lighthill, (1967)}]{whitham_variational_1967}
{\sc {Whitham, G.~B.} \& {Lighthill M.~J.}} 1967 Variational
methods and applications to water waves. {\em
Proc. R. Soc. Lon. Ser. A\/} {\bf
299}~(1456), 6--25.

\bibitem[Zabusky \& Galvin(1971)]{zabusky_shallow-water_1971}
{\sc Zabusky, N.~J. \& Galvin, C.~J.} 1971 Shallow-water waves, the
{Korteweg}-{deVries} equation and solitons. {\em J. Fluid
Mech.\/} {\bf 47}~(4), 811--824.

\end{thebibliography}
\end{document}